\documentclass[journal]{IEEEtran}%
  \usepackage{graphicx}
\usepackage{amsfonts,amsmath}
\usepackage{tablefootnote}
\newtheorem{theorem}{Theorem}

\newtheorem{corollary}[theorem]{Corollary}

\newtheorem{lemma}{Lemma}

\newtheorem{problem}{Problem}

\newtheorem{remark}{Remark}

\begin{document}

\title{Optimal Control of Autonomous Vehicles Approaching A Traffic Light}
\author{Xiangyu~Meng, and Christos~G.~Cassandras \thanks{This work was supported in
part by NSF under grants ECCS-1509084, IIP-1430145, and CNS-1645681, by AFOSR
under grant FA9550-12-1-0113, by DOE under grant DOE-46100, and by Bosch and
the MathWorks.} \thanks{The authors are with the Division of Systems
Engineering and Center for Information and Systems Engineering, Boston
University, Brookline, MA 02446 USA e-mail: \{xymeng,cgc\}@bu.edu.}}
\maketitle

\begin{abstract}
This paper devotes to the development of an optimal acceleration/speed profile
for autonomous vehicles approaching a traffic light. The design objective is
to achieve both short travel time and low energy consumption as well as avoid
idling at a red light. This is achieved by taking full advantage of the
traffic light information based on vehicle-to-infrastructure communication.
The problem is modeled as a mixed integer programming, which is equivalently
transformed into optimal control problems by relaxing the integer constraint.
Then the direct adjoining approach is used to solve both free and fixed
terminal time optimal control problems subject to state constraints. By an
elaborate analysis, we are able to produce a real-time online analytical
solution, distinguishing our method from most existing approaches based on
numerical calculations. Extensive simulations are executed to compare the
performance of autonomous vehicles under the proposed speed profile and human
driving vehicles. The results show quantitatively the advantages of the
proposed algorithm in terms of energy consumption and travel time.

\end{abstract}

\section{Introduction}

The alarming state of existing transportation systems has been well
documented. For instance, in 2014, congestion caused vehicles in urban areas
to spend 6.9 billion additional hours on the road at a cost of an extra 3.1
billion gallons of fuel, resulting in a total cost estimated at \$160 billion
\cite{schrank20152015}. From a control and optimization standpoint, the
challenges stem from requirements for increased safety, increased efficiency
in energy consumption, and lower congestion both in highway and urban
traffic.
Connected and automated vehicles (CAVs), commonly known as self-driving or
autonomous vehicles, provide an intriguing opportunity for enabling users to
better monitor transportation network conditions and to improve traffic flow.
Their proliferation has rapidly grown, largely as a result of Vehicle-to-X (or
V2X) technology \cite{li2014survey} which refers to an intelligent
transportation system where all vehicles and infrastructure components are
interconnected with each other. Such connectivity provides precise knowledge
of the traffic situation across the entire road network, which in turn helps
optimize traffic flows, enhance safety, reduce congestion, and minimize
emissions. Controlling a vehicle to improve energy consumption has been
studied extensively, e.g., see \cite{gilbert1976vehicle, hooker1998optimal,
hellstrom2010design, li2012minimum}. By utilizing road topography information,
an energy-optimal control algorithm for heavy diesel trucks is developed in
\cite{hellstrom2010design}. Based on Vehicle-to-Vehicle (V2V) communication, a
minimum energy control strategy is investigated in car-following scenarios in
\cite{li2012minimum}. Another important line of research focuses on
coordinating vehicles at intersections to increase traffic flow while also
reducing energy consumption. Depending on the control objectives, work in this
area can be classified as dynamically controlling traffic lights
\cite{fleck2016adaptive} and \ as coordinating vehicles
\cite{milanes2010controller},\cite{alonso2011autonomous}%
,\cite{huang2012assessing},\cite{kim2014mpc}. More recently, an optimal
control framework is proposed in \cite{zhang2016optimal} for CAVs to cross one
or two adjacent intersections in an urban area. The state of art and current
trends in the coordination of CAVs is provided in \cite{rios2017survey}.

Our focus in this paper is on an optimal control approach for a single
autonomous vehicle approaching an intersection in terms of energy consumption
and taking advantage of traffic light information. The term \textquotedblleft
ECO-AND\textquotedblright\ short for \textquotedblleft Economical Arrival and
Departure\textquotedblright) is often used to refer to this problem. Its
solution is made possible by vehicle-to-infrastructure (V2I) communication,
which enables a vehicle to automatically receive signals from upcoming traffic
lights before they appear in its visual range. For example, such a V2I
communication system has been launched in Audi cars in Las Vegas by offering a
traffic light timer on their dashboards: as the car approaches an
intersection, a red traffic light symbol and a \textquotedblleft
time-to-go\textquotedblright\ countdown appear in the digital display and
reads how long it will be before the traffic light ahead turns green
\cite{v2i}. Clearly, an autonomous vehicle can take advantage of such
information in order to go beyond current \textquotedblleft
stop-and-go\textquotedblright\ to achieve \textquotedblleft
stop-free\textquotedblright\ driving. Along these lines, the problem of
avoiding red traffic lights is investigated in  \cite{asadi2011predictive,
kamal2013model, mahler2014optimal, wan2016optimal, de2016eco}. The purpose in
\cite{asadi2011predictive} is to track a target speed profile, which is
generated based on the feasibility of avoiding a sequence of red lights. The
approach uses model predictive control based on a receding horizon. The
authors in \cite{kamal2013model} studied an energy-efficient driving strategy
on roads with varying traffic lights and signals at intersections. with the
goal of avoiding a red light instead of following the host vehicle driven by a
human. Avoiding red lights with probabilistic information at multiple
intersections was considered in \cite{mahler2014optimal}, where the time
horizon is discretized and deterministic dynamic programming is utilized to
numerically compute the optimal control input. The work in
\cite{wan2016optimal} devises the optimal speed profile given the feasible
target time, which is within some green light interval. A velocity pruning algorithm is proposed in~\cite{de2016eco} to identify feasible green windows, and a velocity profile is calculated numerically in terms of energy consumption.

Here, we investigate the optimal control problem of autonomous vehicles
approaching a traffic light where the objective function is a weighted sum of
both travel time and energy consumption. The problem is challenging due to the
following reasons. First, finding a feasible green light interval leads to a
Mixed Integer Programming (MIP) problem formulation. In general, solving MIP
problems requires a significant amount of computation, and the optimality of
the solution is not guaranteed due to the non-convexity of the problem
involved with integer variables. The second reason comes from state
constraints related to speed limits. The inclusion of bounds on state
variables poses a significant challenge for most optimization methods. To
overcome the above difficulties, we devise a two-step method. Specifically, we
first address the problem without the traffic light constraint, which means
that the terminal time is free, and the mixed integer constraints are removed.
If the terminal time obtained from the free terminal time optimal control
problem is within some green light interval, then the problem is solved.
However, if the terminal time falls within some red light interval, then the
optimal terminal time could be either the end of the previous green light
interval or the beginning of the next green light interval by using the
monotonicity property of the objective function. Then, we transform the
original problem into a fixed terminal time optimal control problem. We solve
the fixed terminal time optimal control problem with two different terminal
times, and comparing the corresponding performances leads to the optimal
solution of the original problem. All related optimal control problems with
state constraints are solved by using the direct adjoining approach
\cite{hartl1995survey}. The main contributions of our paper are:

\begin{itemize}
\item Instead of solving this problem numerically as in
\cite{asadi2011predictive} and \cite{mahler2014optimal}, an analytical
solution is obtained.

\item For the free terminal time optimal control problem, it is easy to
characterize the type of the optimal acceleration profile.


\item Due to the on-line and real-time nature of the algorithm, the optimal
control profile can be re-calculated as needed, for example when the optimal
trajectory is interrupted by other road users due to safety constraints.
\end{itemize}

The remainder of this paper is organized as follows. The problem is formulated
in Section~\ref{s2}. In Section~\ref{mr}, we present the methodology to solve
the formulated problem, where the solution to the free terminal time optimal
control problem is described in Subsection~\ref{fttp}, and the solution to the
fixed terminal time optimal control problem is presented in
Subsection~\ref{fttoc}. Simulation results illustrating the use of the
proposed algorithm are presented in Section~\ref{ne}. Section~\ref{con}
summarizes our findings, concludes the paper and provides directions for
future work.

\section{Problem Formulation}

\label{s2} 
\begin{figure}[t]
\centering
\includegraphics[width=\linewidth]{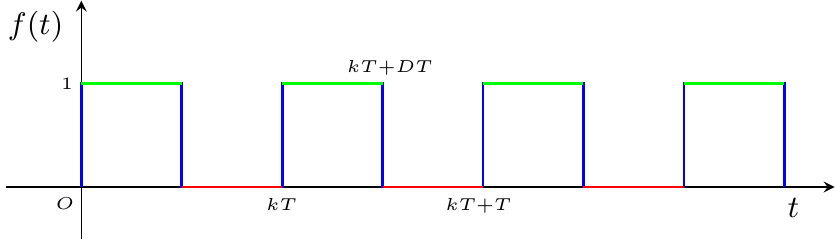}\caption{Traffic light signal}%
\label{fig100}%
\end{figure}

The dynamics of the vehicle are modeled by a double integrator%
\begin{align}
\dot{x}\left(  t\right)    & =v\left(  t\right)  ,\label{dc1}\\
\dot{v}\left(  t\right)    & =u\left(  t\right)  ,\label{dc2}%
\end{align}
where $x\left(  t\right)  ,$ $v\left(  t\right)  $, and $u\left(  t\right)  $
are the position, velocity, and acceleration of the vehicle, respectively. At
time $t_{0},$ the initial position and velocity are given as $x\left(
t_{0}\right)  =0$ and $v\left(  t_{0}\right)  =v_{0}$ respectively. Let us use
$l$ to denote the distance to the traffic light, and $t_{p}$ the intersection
crossing time of the vehicle. The traffic light switches between green and red
at an intersection are dictated by a rectangular pulse signal $f\left(
t\right)  $ with a period $T$:%
\[
f\left(  t\right)  =\left\{
\begin{array}
[c]{ll}%
1 & \text{for }kT\leq t\leq kT+DT,\\
0 & \text{for }kT+DT<t<\left(  k+1\right)  T,
\end{array}
\right.
\]
where $f\left(  t\right)  =1$ indicates that the traffic light is green, and
$f\left(  t\right)  =0$ indicates that the traffic light is red as shown in
Fig.~\ref{fig100}. The parameter $0<D<1$ is the fraction of the time period
$T$ during which the traffic light is green, and $k\in\mathbb{Z}_{\geq0}$ is a
non-negative integer. 

Our objective is to make the vehicle cross an intersection without stopping
with the aid of traffic light information (TLI) as well as to minimize both
travel time and energy consumption. Thus, we formulate the following problem:

\begin{problem}
\label{P1} ECO-AND Problem%
\begin{equation}
\min_{u\left(  t\right)  }\quad\rho_{t}\left(  t_{p}-t_{0}\right)  +\rho
_{u}\int_{t_{0}}^{t_{p}}u^{2}\left(  t\right)  dt\label{OP}%
\end{equation}
subject to%
\begin{align}
&  (\ref{dc1})\text{ and }(\ref{dc2}),\\
&  x\left(  t_{p}\right)  =l,\label{cons1}\\
&  v_{\min}\leq v\left(  t\right)  \leq v_{\max}\label{cons2}\\
&  u_{\min}\leq u\left(  t\right)  \leq u_{\max}\label{cons3}%
\end{align}
and%
\begin{equation}
kT\leq t_{p}\leq kT+DT,\label{cons4}%
\end{equation}
for some $k\in\mathbb{Z}_{\geq0}$.
In (\ref{OP}), the term $J^{t}=t_{p}-t_{0}$ is the travel time while
$J^{u}=\int_{t_{0}}^{t_{p}}u^{2}\left(  t\right)  dt$ captures the energy
consumption; see \cite{malikopoulos2008real}.
\end{problem}

In order to normalize these two terms for the purpose of a well-defined
optimization problem, first note that the maximum possible value of $J^{t}$ is
$l/v_{\min}$. Depending on the relationship between $v_{\min}$, $v_{\max}$,
$u_{\max}$ and $l$, there are two different cases for the maximum possible
value of $J^{u}$. The first case is when the road length is long enough so
that the vehicle can accelerate from $v_{\min}$ to $v_{\max}$ by using the
maximum acceleration $u_{\max}$, i.e., when $l\geq v_{\min}\frac{v_{\max
}-v_{\min}}{u_{\max}}+\frac{1}{2}\frac{\left(  v_{\max}-v_{\min}\right)  ^{2}%
}{u_{\max}}$. In this case,%
\[
J^{u}=\int_{t_{0}}^{t_{p}}u_{\max}^{2}dt=\frac{v_{\max}-v_{\min}}{u_{\max}%
}u_{\max}^{2}=\left(  v_{\max}-v_{\min}\right)  u_{\max}\text{.}%
\]
The second case is when the road length is not long enough for the vehicle to
accelerate to the maximum speed. According to the dynamics (\ref{dc1}) and
(\ref{dc2}), we have%
\[
v_{\min}(t_{p}-t_{0})+\frac{1}{2}u_{\max}(t_{p}-t_{0})^{2}=l.
\]
By solving the above quadratic equation, we are able to get%
\[
t_{p}-t_{0}=\frac{\sqrt{v_{\min}^{2}+2u_{\max}l}-v_{\min}}{u_{\max}}.
\]
Therefore, in this case:%
\[
J^{u}=\int_{t_{0}}^{t_{p}}u_{\max}^{2}dt=\left(  \sqrt{v_{\min}^{2}+2u_{\max
}l}-v_{\min}\right)  u_{\max}.
\]
We can now specify the two weighting parameters $\rho_{t}$ and $\rho_{u}$ as
follows:  $\rho_{t}=\rho\frac{v_{\min}}{l}$ and%
\[
\rho_{u}=\left\{
\begin{array}
[c]{ll}%
\frac{1-\rho}{\left(  v_{\max}-v_{\min}\right)  u_{\max}} &
\begin{array}
[c]{l}%
\text{if }l\geq v_{\min}\frac{v_{\max}-v_{\min}}{u_{\max}}\\
\qquad+\frac{1}{2}\frac{\left(  v_{\max}-v_{\min}\right)  ^{2}}{u_{\max}}%
\end{array}
\\
\frac{1-\rho}{\left(  \sqrt{v_{\min}^{2}+2u_{\max}l}-v_{\min}\right)  u_{\max
}} & \text{otherwise}%
\end{array}
\right.
\]
capturing the normalized trade-off between the travel time and energy
consumption by setting $0\leq\rho\leq1$. When $\rho=0$, the problem reduces to
minimizing the energy consumption only; when $\rho=1$, we seek to minimize the
travel time only.

In (\ref{cons2})-(\ref{cons3}), the parameters $v_{\min}\geq0$ and $v_{\max
}>0$ are the minimum and maximum allowable speeds for road vehicles,
respectively, while the parameters $u_{\min}$ and $u_{\max}$ are the maximum
allowable deceleration and acceleration, respectively. Note that hen $u<0$,
the vehicle decelerates due to braking and when $u>0$ the vehicle accelerates.
Finally, the integer constraint (\ref{cons4}) reflects the requirement that
$t_{p}$ belongs to an interval when the light is green (see Fig.
\ref{fig100}).

\section{Main Results}

\label{mr} Problem \ref{P1} is a Mixed Integer Programming (MIP) problem.
Existing approaches to such problems turn out to be computationally very
demanding for off-line computation, not to mention obtaining analytical
solutions in a real-time on-line context. We propose a two-step approach,
which allows us to efficiently obtain an analytical solution on-line. The
first step is to solve Problem \ref{P1} without the integer constraint
(\ref{cons4}). If the optimal arrival time $t_{p}^{\ast}$ is within some green
light interval, then the problem is solved. However, if%
\[
kT+DT<t_{p}^{\ast}<kT+T,
\]
for some $k$, then we solve Problem \ref{P1} twice with the constraint
(\ref{cons4}) replaced by $t_{p}=kD+DT$ and $t_{p}=kT+T$, respectively. We
compare the performance obtained with different terminal times, and the
solution produced by the one with better performance naturally yields the
optimal solution.

Let us first introduce a lemma, which will be used frequently throughout the
following analysis.

\begin{lemma}
\label{L1} Consider the vehicle's dynamics (\ref{dc1}) and (\ref{dc2}) with
the initial conditions $x_{0}$ and $v_{0}$. If the control input $u\left(
t\right)  =u$ is constant during the time interval $\left[  t_{0}%
,t_{1}\right]  $, then%
\begin{align*}
v\left(  t_{1}\right)   & =v_{0}+u\left(  t_{1}-t_{0}\right)  ,\\
x\left(  t_{1}\right)   & =x_{0}+v_{0}\left(  t_{1}-t_{0}\right)  +\frac{1}%
{2}u\left(  t_{1}-t_{0}\right)  ^{2},\\
J^{u}  & =u^{2}\left(  t_{1}-t_{0}\right)  ;
\end{align*}
If the control input $u\left(  t\right)  =u\left(  t_{1}-t\right)  $ with a
constant $u$, then%
\begin{align*}
v\left(  t_{1}\right)   & =v_{0}+\frac{1}{2}u\left(  t_{1}-t_{0}\right)
^{2},\\
x\left(  t_{1}\right)   & =x_{0}+v_{0}\left(  t_{1}-t_{0}\right)  +\frac{1}%
{3}u\left(  t_{1}-t_{0}\right)  ^{3}\text{,}\\
J^{u}  & =\frac{1}{3}u^{2}\left(  t_{1}-t_{0}\right)  ^{3}.
\end{align*}

\end{lemma}

The proof is given in Appendix~\ref{a10}.

In the following, we first seek the optimal solution to Problem \ref{P1}
without the constraint (\ref{cons4}), which is termed ``free terminal time
optimal control problem''.

\subsection{Free Terminal Time Optimal Control Problem}

\label{fttp} The free terminal time optimal control problem is given below.

\begin{problem}
\label{P2} Free Terminal Time Optimal Control Problem%
\begin{equation}
\min_{u\left(  t\right)  }\rho_{t}\left(  t_{p}-t_{0}\right)  +\rho_{u}%
\int_{t_{0}}^{t_{p}}u^{2}\left(  t\right)  dt\label{ftt1}%
\end{equation}
subject to%
\begin{align}
& (\ref{dc1})\text{ and }(\ref{dc2}),\\
& x\left(  t_{p}\right)  =l ,\label{ftt2}\\
& v_{\min} \leq v\left(  t\right)  \leq v_{\max},\label{ftt3}\\
& u_{\min} \leq u\left(  t\right)  \leq u_{\max},\label{ftt4}%
\end{align}
where $\rho_{t}$ and $\rho_{u}$ are given in Section~\ref{s2}.
\end{problem}

From the objective function (\ref{ftt1}), it can be seen that a minimum energy
consumption solution should avoid braking, that is, $u\left(  t\right)  \geq0$
for $t\in\left[  t_{0},t_{p}\right]  $. We will show this fact in the
following lemma.

\begin{lemma}
\label{l2} The optimal solution $u^{*}(t)$ to Problem~\ref{P2} satisfies
$u^{\ast}\left(  t\right)  \geq0$ for all $t\in\left[  t_{0},t_{p}^{*}\right]
$.
\end{lemma}

The proof is given in Appendix~\ref{a1}.

In addition, it follows from this lemma that whenever $v\left(  \tau\right)
=v_{\max}$ (which may not be possible in some cases), we must have $u\left(
t\right)  =0$ for all $t\in\left[  \tau,t_{p}\right]  $. Based on these
observations, we can derive necessary conditions for the solution to Problem
\ref{P2}, which are summarized in the following theorem.

\begin{theorem}
\label{T1}Let $x^{\ast}\left(  t\right)  $, $v^{\ast}\left(  t\right)  $,
$u^{\ast}\left(  t\right)  $, $t_{p}^{\ast}$ be an optimal solution to Problem
\ref{P2} and assume that $\rho_{t}\neq0$ and $\rho_{u}\neq0$. Then, the
optimal control $u^{\ast}\left(  t\right)  $ satisfies%
\begin{equation}
u^{\ast}\left(  t\right)  =\arg\min_{0\leq u\left(  t\right)  \leq u_{\max}%
}\rho_{u}u^{2}+\frac{\rho_{t}}{v^{\ast}\left(  t_{p}^{\ast}\right)  }\left(
t-\tau\right)  u,\label{NEC1}%
\end{equation}
where $\tau$ is the first time on the optimal path when $v\left(  \tau\right)
=v_{\max}$ if $\tau<t_{p}$; $\tau=t_{p}^{\ast}$ otherwise.
\end{theorem}

\begin{IEEEproof}
Here we use the direct adjoining approach in \cite{hartl1995survey} to
obtain necessary conditions for the optimal solution $u^{\ast }\left(
t\right) $ and $t_{p}^{\ast }$. The Hamiltonian $H\left( v,u,\lambda\right)$ and
Lagrangian $L\left( v,u,\lambda ,\mu ,\eta \right)$ are defined as%
\begin{equation}
H\left( v,u,\lambda\right) =\rho _{u}u^{2}+\rho _{t}+\lambda
_{1}v+\lambda _{2}u  \label{ham}
\end{equation}%
and%
\begin{align}
L\left( v,u,\lambda ,\mu ,\eta \right) =&H\left( v,u,\lambda\right)
+\mu \left( u-u_{\max }\right)  \notag \\
&+\eta _{1}\left( v_{\min }-v\right) +\eta _{2}\left( v-v_{\max }\right) ,
\label{Lag}
\end{align}%
respectively, where $\lambda(t)=[\lambda_{1}(t)\text{ } \lambda_{2}(t)]^{T}$ and
$\eta(t)=[\eta_{1}(t)\text{ } \eta_{2}(t)]^{T}$,%
\begin{align}
&\mu \left( t\right) \geq 0\text{, }\mu \left( t\right) \left[ u^{\ast
}\left( t\right) -u_{\max }\right] =0,  \label{NC1}\\
& \eta _{1}\left( t\right) \geq 0,\text{ }\eta _{2}\left( t\right) \geq 0,
\notag \\
& \eta _{1}\left( t\right) \left[ v_{\min }-v^{\ast }\left( t\right) \right]
+\eta _{2}\left( t\right) \left[ v^{\ast }\left( t\right) -v_{\max }\right]
=0.  \label{NC2}
\end{align}%
Note that we did not include the constraint $u\left( t\right) \geq u_{\min }$
since we have already established that the optimal control $u^{\ast }\left(
t\right) \geq 0$ in the free terminal time optimal control problem in Lemma~\ref{l2}.
Let us temporarily assume that both $\rho _{t}\neq 0$ and $\rho
_{u}\neq 0$. According to Pontryagin's minimum principle, the optimal
control $u^{\ast }\left( t\right) $ must satisfy%
\begin{equation}
u^{\ast }\left( t\right) =\arg \min_{0\leq u\left( t\right) \leq u_{\max
}}H\left( v^{\ast }\left( t\right) ,u(t),\lambda
\left( t\right) \right) \text{,}  \label{PMP}
\end{equation}%
which allows us to express $u^{\ast }\left( t\right) $ in terms of the
costate $\lambda \left( t\right)$, resulting in%
\begin{equation}
u^{\ast }\left( t\right) =\min \left\{ u_{\max },-\frac{\lambda _{2}\left(
t\right) }{2\rho _{u}}\right\},  \label{NCO2}
\end{equation}%
with $\lambda _{2}\left( t\right) \leq 0$ due to Lemma~\ref{l2}. The Lagrange multiplier $\mu
\left( t\right) $ is such that%
\begin{equation}
\left. \frac{\partial L^{\ast }}{\partial u}\right\vert _{u=u^{\ast }\left(
t\right) }=2\rho _{u}u^{\ast }\left( t\right) +\lambda _{2}\left(
t\right) +\mu\left( t\right) =0.  \label{NCO1}
\end{equation}%
Since we can always find $\mu\left( t\right) \geq 0$ to make (\ref%
{NC1}) and (\ref{NCO1}) hold under the minimum principle (\ref{NCO2}),
(\ref{NC1}) and (\ref{NCO1}) can be considered as redundant conditions.
For the costate $\lambda _{1}\left( t\right) $, we have%
\begin{equation*}
\dot{\lambda}_{1}\left( t\right) =-\frac{\partial L^{\ast }\left( t\right) }{%
\partial x}=0,
\end{equation*}%
which means $\lambda _{1}(t)=\lambda _{1}$ is a constant. The costate $\lambda _{2}\left(
t\right) $ satisfies%
\begin{equation}
\dot{\lambda}_{2}\left( t\right) =-\frac{\partial L^{\ast }\left( t\right) }{%
\partial v}=-\lambda _{1}+\eta _{1}\left( t\right) -\eta _{2}\left( t\right) \text{.}  \label{NC5}
\end{equation}
First, let us use a proof by contradiction to show that if $v^{*}\left( t\right)
=v_{\min }$, then $t=t_{0}$. Assume that $v^{*}\left( t\right) =v_{\min }$ for $t\neq
t_{0}$. Then, we must have $v^{*}\left( t\right) =v_{\min }$ for all $t\in \left[
t_{0},t_{p}^{\ast }\right] $. This is because acceleration always precedes cruising at constant speed in the
optimal control profile. If not, the vehicle would travel a longer time for the same trip using the same amount of energy.
According to the system dynamics in (\ref{dc2}), $%
u\left( t\right) =0$ for all $t\in \left[ t_{0},t_{p}^{\ast }\right] $.
Based on the minimum principle (\ref{NCO2}), $\lambda _{2}\left( t\right) =0$
for all $t\in \left[ t_{0},t_{p}^{\ast }\right] $. From (\ref{NC2}), we know
that $\eta _{2}\left( t\right) =0$ for all $t\in \left[
t_{0},t_{p}^{\ast }\right] $. Since the terminal time $t_{p}$ is
unspecified, there is a necessary transversality condition for $t_{p}^{\ast
} $ to be optimal, namely, $H\left( v^{*}(t_{p}^{\ast }),u^{*}(t_{p}^{*}),\lambda(t_{p}^{\ast })\right) =0,$ that
is,%
\begin{equation}
\rho _{u}u^{*}\left( t_{p}^{\ast }\right) ^{2}+\rho _{t}+\lambda _{1}v^{\ast
}\left( t_{p}^{\ast }\right) +\lambda _{2}\left( t_{p}^{\ast }\right)
u\left( t_{p}^{\ast }\right) =0\text{.}  \label{NEC4}
\end{equation}%
Since $u^{*}\left( t_{p}^{\ast }\right) =0,$ we must have $\lambda _{1}<0$
according to (\ref{NEC4}). Then, we obtain $\dot{\lambda}_{2}\left( t\right) > 0$
from (\ref{NC5}), which contradicts $\lambda _{2}\left( t\right) =0$ for $%
t\in \left[ t_{0},t_{p}^{\ast }\right] $. We have thus established that  if $v^{*}\left( t\right)
=v_{\min }$, then $t=t_{0}$.
Next, we will show that $\lambda _{2}\left( t\right) $ has no
discontinuities. Since it is impossible that $v\left( t\right) =v_{\min }$ for $%
t\neq t_{0} $, the costate trajectory $\lambda _{2}\left( t\right) $ may jump
only at some time $\tau $ when $v\left( \tau \right) =v_{\max }$. The condition%
\begin{equation*}
H^{\ast }\left( \tau ^{-}\right) =H^{\ast }\left( \tau ^{+}\right) ,
\end{equation*}%
can be written as%
\begin{eqnarray}
&&\rho _{u}u^{\ast }\left( \tau ^{-}\right) ^{2}+\lambda _{2}\left( \tau
^{-}\right) u^{\ast }\left( \tau ^{-}\right)  \notag \\
&&\qquad =\rho _{u}u^{\ast }\left( \tau ^{+}\right) ^{2}+\lambda _{2}\left(
\tau ^{+}\right) u^{\ast }\left( \tau ^{+}\right) ,  \label{NC11}
\end{eqnarray}%
where $\tau ^{+}$ and $\tau ^{-}$ denote the left-hand side and the
right-hand side limits, respectively. We know from Lemma~\ref{l2}
that $u^{\ast }\left( t\right) =0,$ for $t\in \left[ \tau ,t_{p}^{\ast }%
\right] $. Therefore, from (\ref{NC11}), we obtain%
\begin{equation}
\rho _{u}u^{\ast }\left( \tau ^{-}\right) ^{2}+\lambda _{2}\left( \tau
^{-}\right) u^{\ast }\left( \tau ^{-}\right) =0\text{.}  \label{NC}
\end{equation}%
According to (\ref{NCO2}), we either have $u^{\ast }\left( \tau ^{-}\right)
=-\frac{\lambda _{2}\left( \tau ^{-}\right) }{2\rho _{u}}$ or $u^{\ast
}\left( \tau ^{-}\right) =u_{\max }$. When $u^{\ast }\left( \tau ^{-}\right)
=-\frac{\lambda _{2}\left( \tau ^{-}\right) }{2\rho _{u}},$ (\ref{NC})
becomes%
\begin{equation*}
-\rho _{u}u^{\ast }\left( \tau ^{-}\right) ^{2}=0,
\end{equation*}%
which implies $u^{\ast }\left( \tau ^{-}\right) =\lambda _{2}\left( \tau
^{-}\right) =0$. When $u^{\ast }\left( \tau ^{-}\right) =u_{\max },$ (\ref%
{NC}) becomes%
\begin{equation*}
u_{\max }=\frac{-\lambda _{2}\left( \tau ^{-}\right) }{\rho _{u}},
\end{equation*}%
which contradicts the condition (\ref{NCO2}) where $u_{\max }=\frac{-\lambda
_{2}\left( \tau ^{-}\right) }{2\rho _{u}}$. Therefore, only the case $%
u^{\ast }\left( \tau ^{-}\right) =\lambda _{2}\left( \tau ^{-}\right) =0$ is
possible. In other words, the costate trajectory $\lambda _{2}\left(
t\right) $ has no discontinuities, and the following jump conditions:%
\begin{equation}
\lambda _{2}\left( \tau ^{-}\right) =\lambda _{2}\left( \tau ^{+}\right)
-\zeta _{1}\left( \tau \right) +\zeta _{2}\left( \tau \right) ,  \label{NC9}
\end{equation}%
and%
\begin{align}
&\zeta _{1}\left( \tau \right) \geq 0,\zeta _{2}\left( \tau \right) \geq 0,
\notag \\
&\zeta _{1}\left( \tau \right) \left[ v_{\min }-v^{\ast }\left( \tau \right) %
\right] +\zeta _{2}\left( \tau \right) \left[ v^{\ast }\left( \tau \right)
-v_{\max }\right] =0,  \label{NC10}
\end{align}%
are always satisfied with $\zeta _{1}\left( \tau \right) =\zeta _{2}\left(
\tau \right) =0$.
Next, we will show that $\lambda _{2}\left( t_{p}^{\ast }\right) =0$. At the
terminal time $t_{p}^{*}$, the following transversality conditions hold:%
\begin{eqnarray*}
\lambda _{2}\left( t_{p}^{\ast -}\right)  &=&\gamma _{1}\left. \frac{%
\partial }{\partial v}\left[ v_{\min }-v\right] \right\vert _{v=v^{\ast
}\left( t_{p}^{\ast }\right) } \\
&&+\gamma _{2}\left. \frac{\partial }{\partial v}\left[ v-v_{\max }\right]
\right\vert _{v=v^{\ast }\left( t_{p}^{\ast }\right) }
\end{eqnarray*}%
that is,%
\begin{equation}
\lambda _{2}\left( t_{p}^{*-}\right) =-\gamma _{1}+\gamma _{2},  \label{NC7}
\end{equation}%
where%
\begin{eqnarray}
&&\gamma _{1}\geq 0,\text{ }\gamma _{2}\geq 0,  \notag \\
&&\gamma _{1}\left[ v_{\min }-v^{\ast }\left( t_{p}^{*}\right) \right] +\gamma
_{2}\left[ v^{\ast }\left( t_{p}^{*}\right) -v_{\max }\right] =0.  \label{NC8}
\end{eqnarray}%
If $v_{\min }<v^{*}\left( t_{p}^{\ast }\right) <v_{\max }$, then $\gamma
_{1}=\gamma _{2}=0$, which leads to $\lambda _{2}\left( t_{p}^{\ast }\right)
=0$ by the continuity of $\lambda _{2}\left( t\right) $. When $v^{*}\left(
t_{p}^{\ast }\right) =v_{\max },$ then $u^{*}\left( t_{p}^{\ast }\right) =0$,
which results in $\lambda _{2}\left( t_{p}^{\ast }\right) =0$ according to (%
\ref{NCO2}).
Last, we will show that $\eta _{1}\left( t\right) =0$, and%
\begin{equation*}
\eta _{2}\left( t\right) =\left\{
\begin{array}{cc}
0 & \text{for }t\in \left[ t_{0},\tau \right) \\
-\lambda _{1} & \text{for }t\in \left[ \tau ,t_{p}^{\ast }\right]%
\end{array}%
\right.
\end{equation*}
Since $H\left( v,u,\lambda\right)$ is not an explicit function of time $t$, it follows that%
\begin{equation*}
\frac{dH^{\ast }\left( t\right) }{dt}=0,
\end{equation*}%
that is,%
\begin{equation}
\left[ 2\rho _{u}u^{\ast }\left( t\right) +\lambda _{2}\left(
t\right) \right] \dot{u}^{\ast }\left( t\right) +\left[ \eta _{1}\left( t\right) -\eta _{2}\left( t\right) \right] u^{\ast }\left(
t\right) =0.  \label{TIC}
\end{equation}%
The first term $\left[ 2\rho _{u}u^{\ast }\left( t\right) +\lambda
_{2}\left( t\right) \right] \dot{u}^{\ast }\left( t\right) $ is
always zero since when $u^{\ast }\left( t\right) \neq u_{\max }$, $2\rho
_{u}u^{\ast }\left( t\right) +\lambda _{2}^{\ast }\left( t\right) =0$
according to (\ref{NCO2}), and when $u^{\ast }\left( t\right) =u_{\max }$, $%
\dot{u}^{\ast }\left( t\right) =0$. The condition (\ref{TIC}) can thus be
reduced to%
\begin{equation}
\left[ \eta _{1}\left( t\right) -\eta _{2}\left( t\right) %
\right] u^{\ast }\left( t\right) =0.  \label{NC6}
\end{equation}%
When $v_{0}=v_{\min }$, we have $\eta _{2}\left( t_{0}\right) =0$ from the fact that if $v^{*}\left( t\right)
=v_{\min }$, then $t=t_{0}$ shown earlier and from (\ref%
{NC2}). Condition (\ref{NC6}) then implies%
\begin{equation*}
\eta _{1}\left( t_{0}\right) u^{\ast }\left( t_{0}\right) =0\text{.}
\end{equation*}%
Since $u^{\ast }\left( t_{0}\right) >0$, we can get $\eta _{1}\left(
t_{0}\right) =0$. For $t\neq t_{0},$ $\eta _{1}\left( t\right) =0$ since $v(t)> v_{\min}$ for $t\neq t_{0}$.
Therefore, for any $v_{0},$ we have $\eta _{1}\left( t\right) =0$.
It is easy to get from (\ref{NC2}) that $\eta _{2}\left( t\right) =0$
for $t\in \left[ t_{0},\tau \right) $. For $t\in \left[ \tau ,t_{p}^{\ast }%
\right] $, $\eta _{2}\left( t\right) =-\lambda _{1}$ satisfies the
condition (\ref{NC2}) and $\dot{\lambda}_{2}\left( t\right) =0$ in (\ref{NC5}%
).
Based on the above observations, the differential equation (\ref{NC5})
becomes%
\begin{equation}
\dot{\lambda}_{2}\left( t\right) =-\lambda _{1}  \label{de1}
\end{equation}%
for $t\in \left[ t_{0},\tau \right) $. From (\ref{NEC4}), we have $-\lambda
_{1}=\frac{\rho _{t}}{v^{\ast }\left( t_{p}^{\ast }\right) }$ since $u^{\ast
}\left( t_{p}^{\ast }\right) =0$. Solving the differential equation (\ref%
{de1}), we have%
\begin{equation}
\lambda _{2}\left( t\right) =\frac{\rho _{t}}{v^{\ast }\left( t_{p}^{\ast
}\right) }\left( t-\tau \right)  \label{NEC2}
\end{equation}%
for $t\in \left[ t_{0}\,\tau \right] .$ In the case that $v^{\ast }\left(
t_{p}^{\ast }\right) < v_{\max}$, we simply let $\tau =t_{p}^{\ast }$ in (%
\ref{NEC2}). The proof is completed by substituting (\ref{NEC2}) for $%
\lambda _{2}\left( t\right) $ in (\ref{PMP}).
\end{IEEEproof}

Recall that the theorem was proved under the assumption that $\rho_{t}\neq0$
and $\rho_{u}\neq0$. The special cases when either $\rho_{t}=0$ or $\rho
_{u}=0$ are considered in the following two corollaries.

\begin{corollary}
Let $x^{\ast}\left(  t\right)  $, $v^{\ast}\left(  t\right)  $, $u^{\ast
}\left(  t\right)  $, $t_{p}^{\ast}$ be an optimal solution to Problem
\ref{P2} when $\rho_{t}=0$. Then, the optimal control $u^{\ast}\left(
t\right)  $ satisfies%
\begin{equation}
u^{\ast}\left(  t\right)  =0,\label{NECC1}%
\end{equation}
for all $t\in\lbrack t_{0},t_{p}^{\ast}]$.
\end{corollary}

\begin{corollary}
Let $x^{\ast}\left(  t\right)  $, $v^{\ast}\left(  t\right)  $, $u^{\ast
}\left(  t\right)  $, $t_{p}^{\ast}$ be an optimal solution to Problem
\ref{P2} when $\rho_{u}=0$. Then, the optimal control $u^{\ast}\left(
t\right)  $ satisfies%
\begin{equation}
u^{\ast}\left(  t\right)  =\left\{
\begin{array}
[c]{ll}%
u_{\max} & \text{for }t\in\left[  t_{0},\tau\right)  ,\\
0 & \text{for }t\in\left[  \tau,t_{p}^{\ast}\right]  ,
\end{array}
\right.  \label{NECC2}%
\end{equation}
where $\tau$ is the first time on the optimal path when $v^{\ast}%
(\tau)=v_{\max}$.
\end{corollary}

The proofs of the above two corollaries are straightforward by setting
$\rho_{t}=0$ and $\rho_{u}=0$, respectively, in (\ref{NEC1}) in Theorem
\ref{T1}.

Based on the vehicle dynamics (\ref{dc1}) and (\ref{dc2}), the initial
conditions $x\left(  t_{0}\right)  =0$ and $v\left(  t_{0}\right)  =v_{0},$
and the terminal condition $x^{\ast}\left(  t_{p}^{\ast}\right)  =l$, the
optimal control law (\ref{NEC1}) and the optimal time $t_{p}^{\ast}$ can be
uniquely determined. In the following, we will classify the results into
different cases dependent on the values of the model parameters. In order to
do so, we define two functions:%
\begin{align*}
f\left(  v_{0}\right)    & =l-\frac{v_{\max}^{2}-v_{0}^{2}}{2u_{\max}}%
-u_{\max}v_{\max}^{2}\frac{\rho_{u}}{\rho_{t}}+\frac{1}{6}u_{\max}^{3}v_{\max
}^{2}\frac{\rho_{u}^{2}}{\rho_{t}^{2}},\\
g\left(  v_{0}\right)    & =\,l-2v_{0}\sqrt{\left(  v_{\max}-v_{0}\right)
v_{\max}\frac{\rho_{u}}{\rho_{t}}}\\
& -\frac{4}{3}\left(  v_{\max}-v_{0}\right)  \sqrt{\left(  v_{\max}%
-v_{0}\right)  v_{\max}\frac{\rho_{u}}{\rho_{t}}}.
\end{align*}
Depending on the signs of these two functions, the optimal solution consisting
of $u^{\ast}\left(  t\right)  $ and $t_{p}^{\ast}$ can be classified as shown
in Table~\ref{ta1} with all detailed calculations provided in
Appendix~\ref{a2}. Referring to this table, the optimal control is
parameterized by the following function%
\[
\Phi\left(  t|a,b,c\right)  =\left\{
\begin{array}
[c]{ll}%
u_{\max} & \text{when }t\leq a\\
c(t-b) & \text{when }a<t<b\\
0 & \text{when }t\geq b
\end{array}
\right.
\]
\begin{table*}[ptb]
\caption{Optimal solution classification for Problem \ref{P2}}
\begin{center}
\renewcommand{\arraystretch}{2.0}
\begin{tabular}
[c]{c|c|c|c|c}\hline\hline
& \multicolumn{2}{|c|}{$\frac{v_{0}}{v_{\max}}<1-u_{\max}^{2}\frac{\rho_{u}%
}{\rho_{t}}$} & \multicolumn{2}{|c}{$1-u_{\max}^{2}\frac{\rho_{u}}{\rho_{t}%
}\leq\frac{v_{0}}{v_{\max}}$}\\\hline
& $f\left(  v_{0}\right)  \geq0$ & $f\left(  v_{0}\right)  <0$ & $g\left(
v_{0}\right)  \geq0$ & $g\left(  v_{0}\right)  <0$\\\hline
$u^{\ast}$ & $\Phi\left(  t|t_{1},t_{2},\frac{\rho_{t}}{2\rho_{u}v_{\max}%
}\right)  $ & $\Phi\left( t| t_{3},-,\frac{\rho_{t}}{2\rho_{u}v_{1}}\right)
$%
\tablefootnote{The dash in $\Phi$ means that the variable $t$ cannot reach the upper bound, and therefore that case is inapplicable here. Similar explanations apply to other $\Phi$s defined in Table~\ref{ta1}.} &
$\Phi\left( t| -,t_{4},\frac{\rho_{t}}{2\rho_{u}v_{\max}}\right)  $ &
$\Phi\left( t|-,-,\frac{\rho_{t}}{2\rho_{u}v_{2}}\right)  $\\
$t_{p}^{\ast}$ & $\delta_{1}$ & $\delta_{2}$ & $\delta_{3}$ & $\delta_{4}%
$\\\hline
\end{tabular}
\end{center}
\label{ta1}%
\end{table*}The parameters shown in Table~\ref{ta1} are defined as follows:%
\begin{align*}
t_{1} &  =t_{0}+\frac{\left(  1-u_{\max}^{2}\frac{\rho_{u}}{\rho_{t}}\right)
v_{\max}-v_{0}}{u_{\max}},\text{ }t_{3}=t_{0}+\frac{v_{1}-v_{0}}{u_{\max}},\\
t_{2} &  =t_{1}+2u_{\max}v_{\max}\frac{\rho_{u}}{\rho_{t}},\text{ }t_{4}%
=t_{0}+2\sqrt{\left(  v_{\max}-v_{0}\right)  v_{\max}\frac{\rho_{u}}{\rho_{t}%
}},
\end{align*}
where%
\[
v_{1}=\sqrt{\frac{2u_{\max}l+v_{0}^{2}}{1+\frac{4u_{\max}^{2}}{1-\frac
{\rho_{u}}{\rho_{t}}u_{\max}^{2}}\frac{\rho_{u}}{\rho_{t}}+\frac{8}{3}%
\frac{u_{\max}^{4}}{\left(  1-\frac{\rho_{u}}{\rho_{t}}u_{\max}^{2}\right)
^{2}}\frac{\rho_{u}^{2}}{\rho_{t}^{2}}}}%
\]
and $v_{2}$ is the solution of the following equation:%
\[
l=\frac{2}{3}\left(  v_{0}+2v_{2}\right)  \sqrt{\left(  v_{2}-v_{0}\right)
v_{2}\frac{\rho_{u}}{\rho_{t}}}.
\]
The parameters $\delta_{1},\delta_{2},\delta_{3},\delta_{4}$ specifying in
Table~\ref{ta1} the optimal time $t_{p}^{\ast}$ when the vehicle arrives at
the traffic light  in each of the four possible cases are given below:
\begin{align*}
\delta_{1} &  =t_{2}+\frac{f\left(  v_{0}\right)  }{v_{\max}},\\
\delta_{2} &  =t_{3}+2u_{\max}\frac{v_{1}}{1-\frac{\rho_{u}}{\rho_{t}}u_{\max
}^{2}}\frac{\rho_{u}}{\rho_{t}},\\
\delta_{3} &  =t_{4}+\frac{g(v_{0})}{v_{\max}},\\
\delta_{4} &  =t_{0}+2\sqrt{\left(  v_{2}-v_{0}\right)  v_{2}\frac{\rho_{u}%
}{\rho_{t}}}.
\end{align*}

\begin{remark}
This remark pertains to the underlying criteria for the optimal solution
classification in Table~\ref{ta1}. The first row determines whether or not the
maximum acceleration $u_{\max}$ will be used for a given initial speed $v_{0}%
$. The optimality conditions tell us that the vehicle starts with the maximum
acceleration when the initial speed is relatively slow. The second row
determines if the road length $l$ is large enough for a vehicle to reach its
maximum speed for a given initial speed $v_{0}$. In general, the optimal
control contains three phases: full acceleration, linearly decreasing
acceleration, and no acceleration. The first column specifies the case where
all three phases are included with switches defined by $t_{1}$, $t_{2}$. The
second column corresponds to the case of low initial speeds and short-length
roads. Under optimal control in this case, the vehicle starts with full
acceleration, but the road length is so short that the maximum speed cannot be
reached. Therefore, the optimal control contains only the first two phases.
The third column corresponds to the case of large initial speeds and
long-length roads. The vehicle starts with linearly decreasing acceleration,
and then proceeds with no acceleration when the speed reaches the limit
$v_{\max}$. Here, the optimal control contains only the last two phases. The
last column corresponds to the case of large initial speeds and short-length
roads. Therefore, the vehicle uses only linearly decreasing acceleration.
\end{remark}

\subsection{Fixed Terminal Time Optimal Control Problem}

\label{fttoc}

In this section, we consider the case where the optimal time $t_{p}^{\ast}$
obtained in the free terminal time optimal control problem \ref{P2} is within
some red light interval, that is,%
\[
kT+DT<t_{p}^{\ast}<kT+T,
\]
In this case, the candidate optimal arrival time $t_{p}^{\ast}$ in
Problem~\ref{P1} is either $kT+DT$ or $kT+T$. Therefore, we can compare the
performance obtained under either one of these two terminal times, and select
the one with better performance to determine the optimal arrival time for
Problem~\ref{P1}. In both cases, the travel time is now fixed, hence the only
objective is to minimize the energy consumption. Thus, we have the following
problem formulation:

\begin{problem}
\label{P3}Fixed Terminal Time Optimal Control Problem%
\begin{equation}
\min_{u\left(  t\right)  }\int_{t_{0}}^{t_{p}}u^{2}\left(  t\right)
dt\label{ft1}%
\end{equation}
subject to%
\begin{align}
& (\ref{dc1}) \text{ and } (\ref{dc2})\\
& x\left(  t_{p}\right)  =l\label{ft2}\\
& t_{p} =kT+DT\text{ or }kT+T\label{ft3}\\
& v_{\min} \leq v\left(  t\right)  \leq v_{\max}\label{ft4}\\
& u_{\min} \leq u\left(  t\right)  \leq u_{\max}\label{ft5}%
\end{align}

\end{problem}

\subsubsection{Arrival Time $t_{p}=kT+DT$}

In this case, it is clear that that the vehicle must use less time than the
one specified by $t_{p}^{\ast}$ in Problem \ref{P2} and higher acceleration.
Define a function%
\[
h\left(  v_{0}\right)  =\left\{
\begin{array}
[c]{ll}%
v_{0}t_{p}+\frac{1}{2}u_{\max}t_{p}^{2}-l & \text{for }t_{p}\leq\frac{v_{\max
}-v_{0}}{u_{\max}}\\
v_{\max}t_{p}-\frac{1}{2}\frac{\left(  v_{\max}-v_{0}\right)  ^{2}}{u_{\max}%
}-l & \text{for }t_{p}>\frac{v_{\max}-v_{0}}{u_{\max}}%
\end{array}
\right.
\]
Observe that the terminal time $t_{p}=kT+DT$ is possible if and only if
$h\left(  v_{0}\right)  \geq0$. The main result for this case is given in the
following theorem.

\begin{theorem}
Let $x^{\ast}\left(  t\right)  $, $v^{\ast}\left(  t\right)  $, $u^{\ast
}\left(  t\right)  $ be an optimal solution to Problem \ref{P3} with
$t_{p}=kT+DT$. Then, the optimal control $u^{\ast}\left(  t\right)  $
satisfies%
\[
u^{\ast}\left(  t\right)  =\arg\min_{0\leq u\left(  t\right)  \leq u_{\max}%
}u^{2}+\frac{u^{\ast}\left(  t_{0}\right)  ^{2}\left(  t-\tau\right)  u}%
{v_{0}-v^{\ast}\left(  t_{p}\right)  +\left(  \tau-t_{0}\right)  u^{\ast
}\left(  t_{0}\right)  }\text{,}%
\]
where $\tau$ is the first time on the optimal path when $v\left(  \tau\right)
=v_{\max}$ if $\tau<t_{p}$; $\tau=t_{p}^{\ast}$ otherwise.
\end{theorem}

\begin{IEEEproof}
Similar to the proof of Theorem~\ref{T1}, we will use the direct adjoining approach \cite{hartl1995survey} to solve the fixed terminal time optimal control problem.
The Hamiltonian $H\left( v,u,\lambda\right)$ and
Lagrangian $L\left( v,u,\lambda ,\mu ,\eta \right)$ are defined as%
\begin{equation*}
H(u,v,\lambda)=u^{2}+\lambda _{1}v+\lambda _{2}u,
\end{equation*}%
and
\begin{align*}
L(u,v,\lambda,\mu,\eta)=&\,H+\mu \left( u-u_{\max }\right) \\
&+\eta _{1}\left( v_{\min }-v\right) +\eta
_{2}\left( v-v_{\max }\right) ,
\end{align*}%
respectively, where $\lambda(t)=[\lambda_{1}(t)\text{ } \lambda_{2}(t)]^{T}$ and
$\eta(t)=[\eta_{1}(t)\text{ } \eta_{2}(t)]^{T}$, and %
\begin{align*}
&\mu \left( t\right) \geq 0,\mu \left( t\right) \left[ u^{\ast }\left(
t\right) -u_{\max }\right] =0,\\
&\eta _{1}\left( t\right) \geq 0,\eta _{2}\left( t\right) \geq 0,\\
&\eta
_{1}\left( t\right) \left[ v_{\min }-v^{\ast }\left( t\right) \right] +\eta
_{2}\left( t\right) \left[ v^{\ast }\left( t\right) -v_{\max }\right] =0.
\end{align*}%
Note that, as in the the proof of Theorem~\ref{T1}, $u^{\ast }\left( t\right) \geq 0$ for all $t$,
therefore, the constraint $u\left( t\right) \geq u_{\min }$ is relaxed, and%
\begin{equation}
u^{\ast }\left( t\right) =\arg \min_{0\leq u\left( t\right) \leq u_{\max
}}u^{2}+\lambda _{2}u\text{.}\label{op11}
\end{equation}%
which implies that%
\begin{equation*}
u^{\ast }\left( t\right) =\min \left\{ u_{\max },-\frac{\lambda _{2}\left(
t\right) }{2}\right\} \text{,}
\end{equation*}%
and $\lambda _{2}\left( t\right) \leq 0$.
From the proof of Theorem~\ref{T1}, we know that $\mu \left( t\right) $ is a redundant
variable, and $\lambda _{1}$ is a constant. Let us first assume that%
\[
l>v_{\min}t_{p}.
\]
Note that the case of $l=v_{\min}t_{p}$ cannot occur when $t_{p}=kT+DT$ (however, it may occur when $t_{p}=kT+T$ and this case will be discussed later).
Again, we can prove the fact that $v\left( t\right) =v_{\min }$ happens only at $t=t_{0}$ but without using the transversality condition as we did in the free terminal time optimal control problem.
The property that $\lambda _{2}\left( t\right) $ has no discontinuities also still holds. The costate $\lambda_{2}(t)$ satisfies%
\begin{equation}
\dot{\lambda}_{2}\left( t\right) =-\lambda _{1}+\eta _{1}\left(
t\right) -\eta _{2}\left( t\right) \text{.} \label{exp1}
\end{equation}%
Similarly, we
can show that $\eta _{1}\left( t\right) =0$, and (\ref{exp1}) reduces to%
\begin{equation}
\dot{\lambda}_{2}\left( t\right) =-\lambda _{1} \label{exp2}
\end{equation}%
for $t\in \left[ t_{0},\tau \right) $ and $\lambda _{2}\left( \tau \right)
=0 $. By solving the differential equation (\ref{exp2}), we get%
\begin{equation}
\lambda _{2}\left( t\right) =-\lambda _{1}\left( t-\tau \right) . \label{lam2}
\end{equation}%
Again since the Hamiltonian is not an explicit function of time, by the condition%
\begin{equation*}
H\left( t_{0}\right) =H\left( t_{p}\right) \text{,}
\end{equation*}%
we have%
\begin{equation}
u^{\ast }\left( t_{0}\right) ^{2}+\lambda _{1}v^{\ast }\left( t_{0}\right)
-\lambda _{1}\left( t_{0}-\tau \right) u^{\ast }\left( t_{0}\right) =\lambda
_{1}v^{\ast }\left( t_{p}\right)  \label{lamba1}
\end{equation}%
where the fact that $\lambda_{2}(t_{p})=u^{*}(t_{p})=0$ has been used. From (\ref{lamba1}), we can obtain%
\begin{equation}
\lambda _{1}=\frac{u^{\ast }\left( t_{0}\right) ^{2}}{v^{\ast }\left(
t_{p}\right) +\left( t_{0}-\tau \right) u^{\ast }\left( t_{p}\right)
-v^{\ast }\left( t_{0}\right) }\label{lam1}
\end{equation}
For $t\in [\tau, t_{p}]$, we can just let $\eta_{2}(t)=-\lambda_{1}$.
If $v^{*}(t_{p})<v_{\max}$, then $\tau=t_{p}$ in (\ref{lamba1}).
The proof is completed by substituting $\lambda _{1}$ in (\ref{lam1}) into (\ref{lam2}), and then $\lambda _{2}$ into
(\ref{op11}).
\end{IEEEproof}

Given the terminal time $kT+DT$ and the road length $l$, the value of $v_{0}$
can be classified into one of five cases as shown in Table~\ref{ta2}. Note
that if Case $i$ is infeasible for some $v_{0}$ and the given parameters, we
can treat $J_{i}^{u}$ as infinity. \begin{table}[pbh]
\caption{Optimal solution classification for Problem \ref{P3} with
$t_{p}=kT+DT$}%
\renewcommand{\arraystretch}{2.0}
\par
\begin{center}%
\begin{tabular}
[c]{c|cc}\hline
& Optimal Control & Performance\\\hline
Case I & $u_{0}^{*}=u_{\max}$ and $\dot{u}^{*}\left(  t\right)  =0$ &
$J_{1}^{u}$\\
Case II & $u^{\ast}\left(  t_{0}\right)  =u_{\max}$ and $v^{\ast}\left(
t_{p}\right)  =v_{\max}$ & $J_{2}^{u}$\\
Case III & $u^{\ast}\left(  t_{0}\right)  =u_{\max}$ and $v^{\ast}\left(
t_{p}\right)  <v_{\max}$ & $J_{3}^{u}$\\
Case IV & $u_{0}^{\ast}<u_{\max}$ and $v^{\ast}\left(  t_{p}\right)  =v_{\max
}$ & $J_{4}^{u}$\\
Case V & $u_{0}^{\ast}<u_{\max}$ and $v^{\ast}\left(  t_{p}\right)  <v_{\max}$
& $J_{5}^{u}$\\\hline
\end{tabular}
\end{center}
\label{ta2}%
\end{table}The performances associated with each case in Table~\ref{ta2} as
well as the detailed calculations are given in Appendix~\ref{a3}. After
obtaining the performance for each cases with $t_{p}=kT+DT$, we select the one
with the smallest energy consumption, that is,%
\[
J_{u}^{kT+DT}=\min\left\{  J_{1}^{u},\ldots J_{5}^{u}\right\}  ,
\]
with the corresponding optimal acceleration profile.

\subsubsection{Arrival Time $t_{p}=kT+T$}

In this case, the vehicle must use less acceleration than in the free terminal
time case. Depending on the initial speed $v_{0}$, there are three cases to
consider. First, if%
\[
l=v_{0}\left(  kT+T-t_{0}\right)  ,
\]
then the vehicle can cruise through the intersection with the constant speed
$v_{0}$ without any acceleration (Case VI in Table~\ref{ta3}). The energy
consumption in this case is%
\[
J_{6}^{u}=0\text{.}%
\]
If, on the other hand,
\[
l>v_{0}\left(  kT+T-t_{0}\right)  ,
\]
then the problem can be solved using the result of the case $t_{p}=kT+DT$
analyzed above. Finally, if%
\[
l<v_{0}\left(  kT+T-t_{0}\right)  \text{,}%
\]
then the vehicle must decelerate to reach the traffic light while in its green
state. Therefore, the control input is only subject to the constraint%
\[
u_{\min}\leq u\left(  t\right)  \leq0\text{.}%
\]
The main result in this case is given in the following theorem.

\begin{theorem}
Let $x^{\ast}\left(  t\right)  $, $v^{\ast}\left(  t\right)  $, $u^{\ast
}\left(  t\right)  $ be an optimal solution to Problem \ref{P3} with
$t_{p}=KT+T$. Then, the optimal solution $u^{\ast}\left(  t\right)  $
satisfies%
\[
u^{\ast}\left(  t\right)  =\arg\min_{u_{\min}\leq u\left(  t\right)  \leq
0}u^{2}+\frac{u^{\ast}\left(  t_{0}\right)  ^{2}\left(  \tau-t\right)
u}{v^{\ast}\left(  t_{p}\right)  -v_{0}-\left(  \tau-t_{0}\right)  u^{\ast
}\left(  t_{0}\right)  },
\]
where $\tau$ is the first time on the optimal path when $v\left(  \tau\right)
=v_{\max}$ if $\tau<t_{p}$; $\tau=t_{p}^{\ast}$ otherwise.
\end{theorem}

\begin{IEEEproof}
The Hamiltonian $H\left(  u,v,\lambda\right)  $ and the Lagrangian $L\left(
u,v,\lambda,\mu,\eta\right)  $ are defined as%
\[
H\left(  u,v,\lambda\right)  =u^{2}+\lambda_{1}v+\lambda_{2}u
\]
and%
\begin{align*}
L\left(  u,v,\lambda,\mu,\eta\right)    & =H\left(  u,v,\lambda\right)
+\mu\left(  u_{\min}-u\right)  \\
& +\eta_{1}\left(  v_{\min}-v\right)  +\eta_{2}\left(  v-v_{\max}\right)
\text{,}%
\end{align*}
respectively, where $\lambda(t)=\left[  \lambda_{1}(t),\lambda_{2}(t)\right]
^{T}$, $\eta(t)=\left[  \eta_{1}(t),\eta_{2}(t)\right]  ^{T}$, and%
\begin{align*}
&  \mu\left(  t\right)  \geq0,\text{ }\mu\left(  t\right)  \left[  u_{\min
}-u^{\ast}\left(  t\right)  \right]  =0,\\
&  \eta_{1}\left(  t\right)  \geq0,\text{ }\eta_{2}\left(  t\right)  \geq0,\\
&  \eta_{1}\left(  t\right)  \left[  v_{\min}-v^{\ast}\left(  t\right)
\right]  +\eta_{2}\left(  t\right)  \left[  v^{\ast}\left(  t\right)
-v_{\max}\right]  =0.
\end{align*}
As before, we do not include the constraint $u\left(  t\right)  \leq u_{\max}$
since we have already established in Lemma \ref{l2} that $u^{\ast}\left(
t\right)  \leq0$.

According to Pontryagin's minimum principle, the optimal control $u^{\ast
}\left(  t\right)  $ must satisfy%
\[
u^{\ast}\left(  t\right)  =\arg\min_{u_{\min}\leq u\left(  t\right)  \leq
0}H\left(  v^{\ast}\left(  t\right)  ,u^{\ast}\left(  t\right)  ,\lambda
\left(  t\right)  \right)
\]
which allows us to express $u^{\ast}\left(  t\right)  $ in terms of the
costate $\lambda\left(  t\right)  $, that is,%
\begin{equation}
u^{\ast}\left(  t\right)  =\max\left\{  u_{\min},-\frac{\lambda_{2}\left(
t\right)  }{2}\right\} \label{MP}%
\end{equation}
with $\lambda_{2}\left(  t\right)  \geq0$. The Lagrange multiplier $\mu\left(
t\right)  $ is redundant as before. The costate $\lambda_{1}$ is a constant.
The co-state $\lambda_{2}\left(  t\right)  $ satisfies%
\[
\dot{\lambda}_{2}\left(  t\right)  =-\frac{\partial L^{\ast}}{\partial
v}=-\lambda_{1}+\eta_{1}^{\ast}\left(  t\right)  -\eta_{2}^{\ast}\left(
t\right)  \text{.}%
\]

First, it is easy to see that $v_{0}\neq v_{\min}$. Let $\tau$ be the first
time that $v\left(  \tau\right)  =v_{\min}$, then%
\[
u^{\ast}\left(  t\right)  =0
\]
for $t\geq\tau$. Again, since the Hamiltonian is not an explicit function of
time, by the condition%
\[
H^{\ast}\left(  \tau^{-}\right)  =H^{\ast}\left(  \tau^{+}\right)  ,
\]
we have%
\begin{equation}
u^{\ast}\left(  \tau^{-}\right)  ^{2}+\lambda_{2}\left(  \tau^{-}\right)
u^{\ast}\left(  \tau^{-}\right)  =0\text{.}\label{eq1}%
\end{equation}
According to (\ref{MP}), we either have $u^{\ast}\left(  \tau^{-}\right)
=u_{\min}$ or $u^{\ast}\left(  \tau^{-}\right)  =-\frac{\lambda_{2}\left(
\tau^{-}\right)  }{2}$. When $u^{\ast}\left(  \tau^{-}\right)  =u_{\min}$, the
above equality becomes%
\[
u_{\min}^{2}+\lambda_{2}\left(  \tau^{-}\right)  u_{\min}=0\text{,}%
\]
which contradicts the minimum principle (\ref{MP}); when $u^{\ast}\left(
\tau^{-}\right)  =-\frac{\lambda_{2}\left(  \tau^{-}\right)  }{2}$,
(\ref{eq1}) becomes%
\[
u^{2}\left(  \tau^{-}\right)  -2u^{2}\left(  \tau^{-}\right)  =0.
\]
Therefore, only $\lambda_{2}\left(  \tau^{-}\right)  =u^{\ast}\left(  \tau
^{-}\right)  =0$ is possible, that is to say, $\lambda_{2}$ and $u^{\ast}$
have no discontinuities at $\tau$.

At the terminal time $t_{p}$, the following costate boundary condition holds:%
\begin{align*}
\lambda_{2}\left(  t_{p}^{\ast-}\right)    & =\gamma_{1}\left.  \frac
{\partial}{\partial v}\left[  v_{\min}-v\right]  \right\vert _{v=v^{\ast
}\left(  t_{p}^{\ast}\right)  }\\
& +\gamma_{2}\left.  \frac{\partial}{\partial v}\left[  v-v_{\max}\right]
\right\vert _{v=v^{\ast}\left(  t_{p}^{\ast}\right)  }%
\end{align*}
that is,%
\[
\lambda_{2}\left(  t_{p}^{-}\right)  =-\gamma_{1}+\gamma_{2}%
\]
and%
\[
\gamma_{1}\geq0,\text{ }\gamma_{2}\geq0,\text{ }\gamma_{1}\left[  v_{\min
}-v^{\ast}\left(  t_{p}\right)  \right]  +\gamma_{2}\left[  v^{\ast}\left(
t_{p}\right)  -v_{\max}\right]  =0\text{.}%
\]
At $t_{p}$, we know that $v^{\ast}\left(  t_{p}\right)  \neq v_{\max}$. Thus,
$\gamma_{2}=0$. Likewise, it is easy to obtain $\gamma_{1}=0$. Therefore, we
have%
\[
\lambda_{2}\left(  t_{p}\right)  =0\text{.}%
\]

Since the Hamiltonian is not an explicit function of time, the condition%
\[
\frac{dH^{\ast}\left(  t\right)  }{dt}=0,
\]
implies that%
\[
\left[  2u^{\ast}\left(  t\right)  +\lambda_{2}\left(  t\right)  \right]
\dot{u}^{\ast}\left(  t\right)  +\left[  \eta_{1}\left(  t\right)  -\eta
_{2}\left(  t\right)  \right]  u^{\ast}\left(  t\right)  =0\text{.}%
\]
Since the first term is always zero as before, the above condition becomes%
\[
\left[  \eta_{1}\left(  t\right)  -\eta_{2}\left(  t\right)  \right]  u^{\ast
}\left(  t\right)  =0\text{.}%
\]
When $v_{0}=v_{\max}$, we have $\eta_{1}\left(  t_{0}\right)  =0,$ that is%
\[
\eta_{2}\left(  t_{0}\right)  u^{\ast}\left(  t_{0}\right)  =0\text{.}%
\]
Recall that%
\[
\dot{\lambda}_{2}\left(  t\right)  =-\frac{\partial L^{\ast}}{\partial
v}=-\lambda_{1}+\eta_{1}\left(  t\right)  -\eta_{2}\left(  t\right)  \text{.}%
\]
Since $\lambda_{1}>0$, then $\lambda_{2}\left(  t\right)  $ must decrease.
Therefore, $u^{\ast}\left(  t_{0}\right)  <0$, and $\eta_{2}\left(  t\right)
=0$ for all $t$. For $t\in\left[  t_{0},\tau\right)  $, $\eta_{1}^{\ast
}\left(  t\right)  =0$. Therefore,%
\[
\dot{\lambda}_{2}\left(  t\right)  =-\lambda_{1}%
\]
for $t\in\left[  t_{0},\tau\right)  $. For $t\in\left[  \tau,t_{p}\right)  ,$%
\[
\dot{\lambda}_{2}\left(  t\right)  =-\lambda_{1}+\eta_{1}^{\ast}\left(
t\right)  =0.
\]
Solving the above differential equation, we obtain%
\begin{equation}
\lambda_{2}\left(  t\right)  =\lambda_{1}\left(  \tau-t\right)  ,\label{exp3}%
\end{equation}
for $t\in\lbrack t_{0},\tau)$. By the condition%
\[
H\left(  t_{0}\right)  =H\left(  t_{p}\right)  \text{,}%
\]
we have%
\[
u^{\ast}\left(  t_{0}\right)  ^{2}+\lambda_{1}v_{0}+\lambda_{1}\left(
\tau-t_{0}\right)  u^{\ast}\left(  t_{0}\right)  =\lambda_{1}v^{\ast}\left(
t_{p}\right)  \text{,}%
\]
that is,%
\[
\lambda_{1}=\frac{u^{\ast}\left(  t_{0}\right)  ^{2}}{v^{\ast}\left(
t_{p}\right)  -v_{0}-\left(  \tau-t_{0}\right)  u^{\ast}\left(  t_{0}\right)
}.
\]
The proof is completed by substituting $\lambda_{1}$ into (\ref{exp3}) and
then $\lambda_{2}$ into (\ref{MP}).
\end{IEEEproof}

\begin{table}[pbh]
\caption{Optimal solution classification for Problem \ref{P3} with
$t_{p}=kT+T$}%
\renewcommand{\arraystretch}{2.0}
\par
\begin{center}%
\begin{tabular}
[c]{c|cc}\hline
& Optimal Control & Performance\\\hline
Case VI & $u^{*}(t)=0$ and $v^{*}(t)=v_{0}$ & $J_{6}^{u}$\\
Case VII & $u^{*}\left(  t_{0}\right)  =u_{\min}$ and $v^{*}\left(
t_{p}\right)  =v_{\min}$ & $J_{7}^{u}$\\
Case VIII & $u^{\ast}\left(  t_{0}\right)  =u_{\min}$ and $v^{\ast}\left(
t_{p}\right)  >v_{\min}$ & $J_{8}^{u}$\\
Case IX & $u^{\ast}\left(  t_{0}\right)  <u_{\min}$ and $v^{\ast}\left(
t_{p}\right)  =v_{\min}$ & $J_{9}^{u}$\\
Case X & $u^{\ast}\left(  t_{0}\right)  <u_{\min}$ and $v^{\ast}\left(
t_{p}\right)  <v_{\min}$ & $J_{10}^{u}$\\\hline
\end{tabular}
\end{center}
\label{ta3}%
\end{table}The classification of all possible solutions with $t_{p}=kT+T$ is
shown in Table~\ref{ta3}. The performances associated with each case in this
table as well as the detailed calculations are given in Appendix~\ref{a4}.
After obtaining the energy consumption from $J_{6}^{u}$ through $J_{10}^{u}$,
we can select%
\[
J_{u}^{kT+T}=\min\left\{  J_{6}^{u},\ldots,J_{10}^{u}\right\}  \text{,}%
\]
where $J_{i}^{u}$ can be treated as infinity if Case $i$ is infeasible.
Finally, we can compare the two performances obtained, that is,%
\begin{align*}
J^{kT+DT}  & =\rho_{t}\left(  kT+DT\right)  +\rho_{u}J_{u}^{kT+DT}\\
J^{kT+T}  & =\rho_{t}\left(  kT+T\right)  +\rho_{u}J_{u}^{kT+T}%
\end{align*}
and determine the optimal performance to be the one with a smaller value.

\section{Numerical Examples}

\label{ne} 
\begin{figure}[t]
\centering
\includegraphics[width=\linewidth]{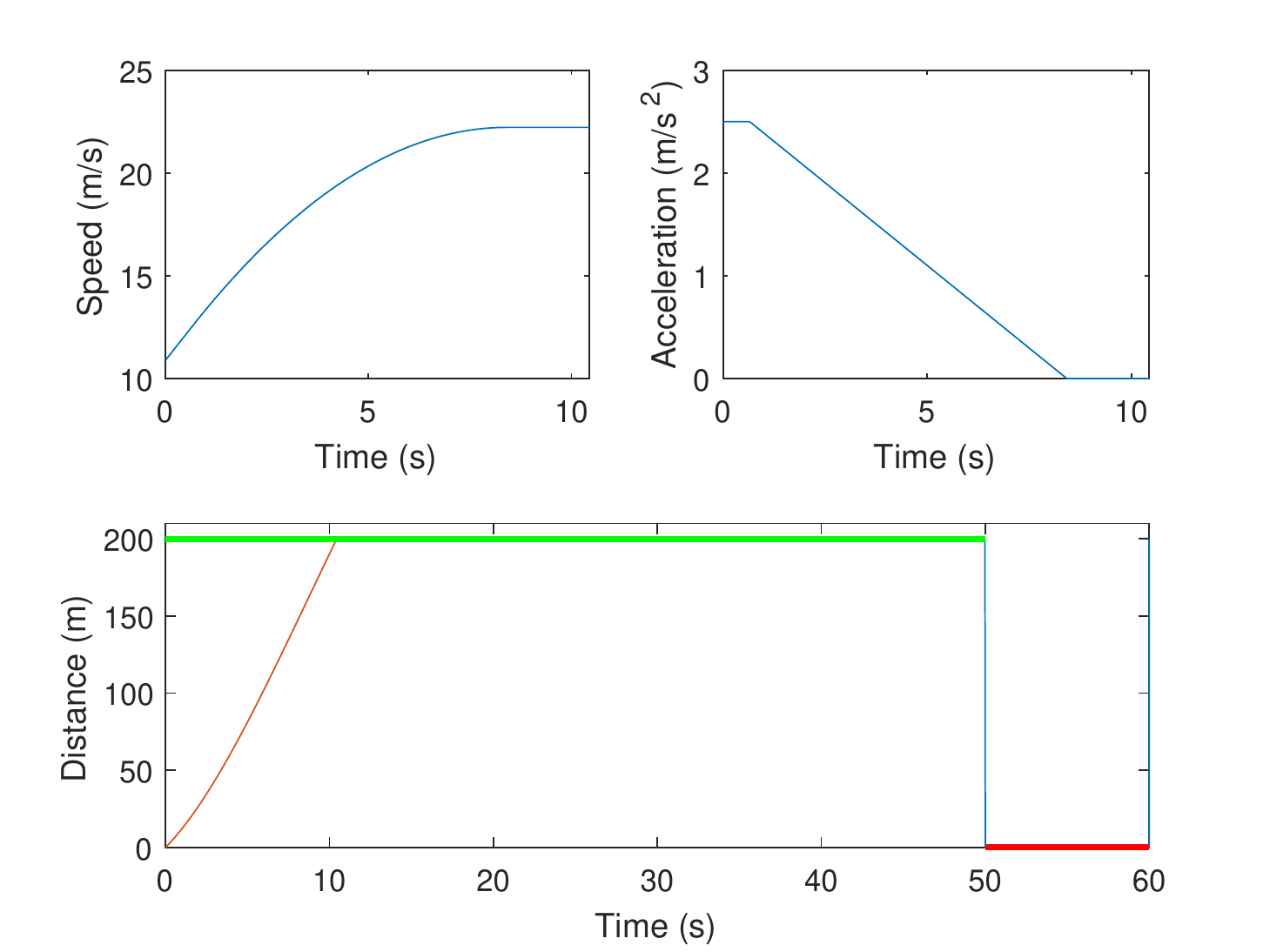}\caption{Case I in Table~\ref{ta1}
with $v_{0}=10.8869$.}%
\label{fig1}%
\end{figure}

\begin{figure}[t]
\centering
\includegraphics[width=\linewidth]{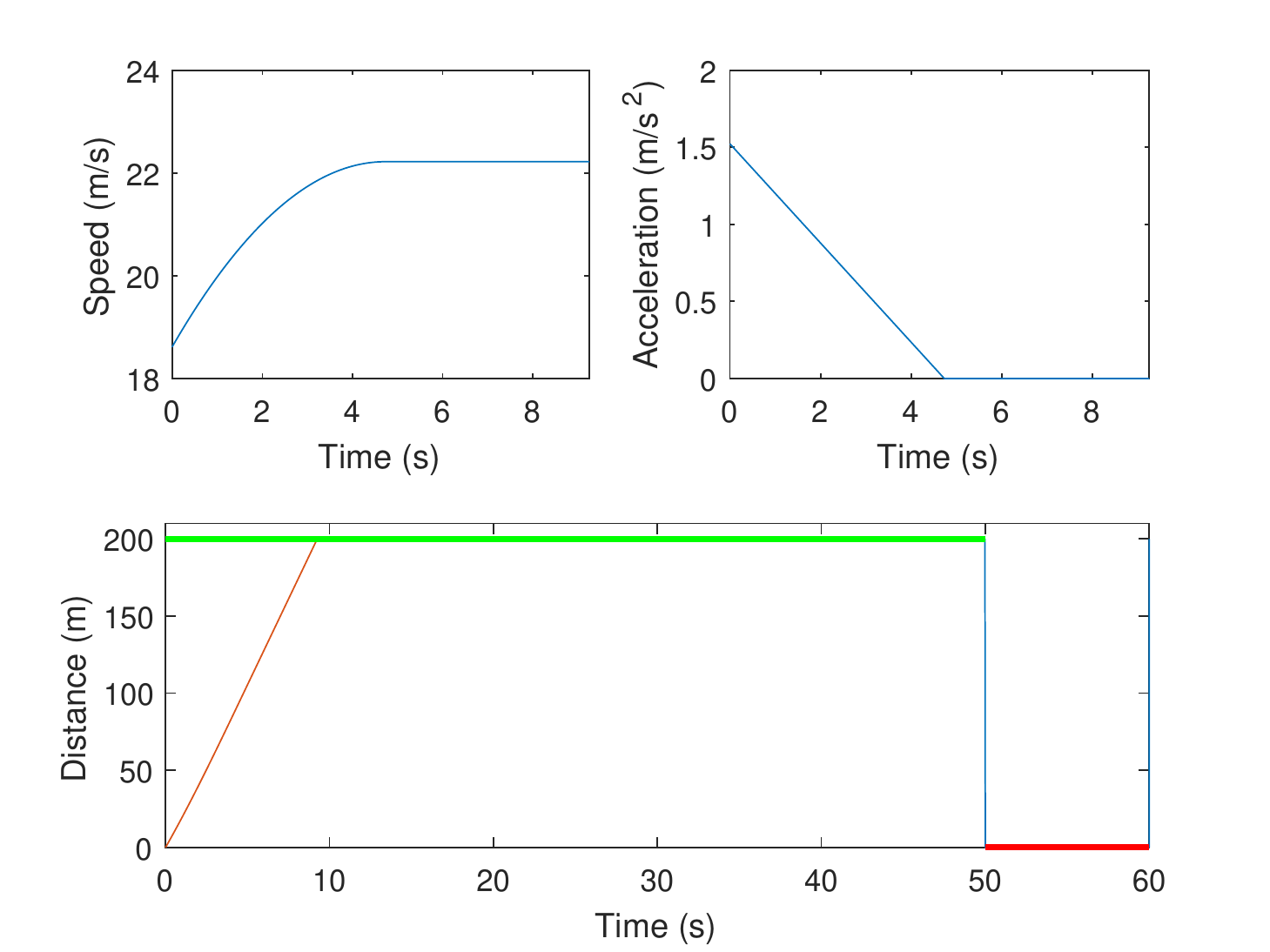}\caption{Case III in
Table~\ref{ta1} with $v_{0}=18.6182$.}%
\label{fig2}%
\end{figure}

We have simulated the system defined by the vehicle dynamics
(\ref{dc1}) and (\ref{dc2}) and associated constraints and optimal control
problem parameters with values given as follows. The minimum and maximum
speeds are 2.78~$m/s$ and 22.22~$m/s$. The maximum acceleration and
deceleration are set to $2.5~m/s^{2}$ and $-2.9~m/s^{2}$, respectively. The
weights in (\ref{OP}) are set using $\rho=0.9549$, that is, $\rho_{t}=0.0133$,
and $\rho_{u}=9.2798\times10^{-4}$. In this case, the values%
\[
1-u_{\max}^{2}\frac{\rho_{u}}{\rho_{t}}=0.5630,
\]
and%
\[
\frac{v_{m}+v_{M}}{2v_{M}}=0.5626,
\]
are almost the same. Thus, if we randomly generate the initial speed $v_{0}$
from a uniform distribution on the interval $[v_{\min},v_{\max}]$, different
initial speeds fall roughly equally into the two different cases in the first
row in Table \ref{ta1}. The total cycle time for the traffic light is 60~$s$
with different patterns. We first test the optimal controller on a road of
length 200~$m$. Figure~\ref{fig1} depicts the case when the initial speed is
relatively slow. The vehicle starts with full acceleration and, when the speed
limit is reached, it switches to no acceleration. The vehicle arrives at the
traffic light within the first green light cycle. When the initial speed is relatively large, the vehicle should not
start with full acceleration. This is the case shown in Fig.~\ref{fig2}.

\begin{figure}[t]
\centering
\includegraphics[width=\linewidth]{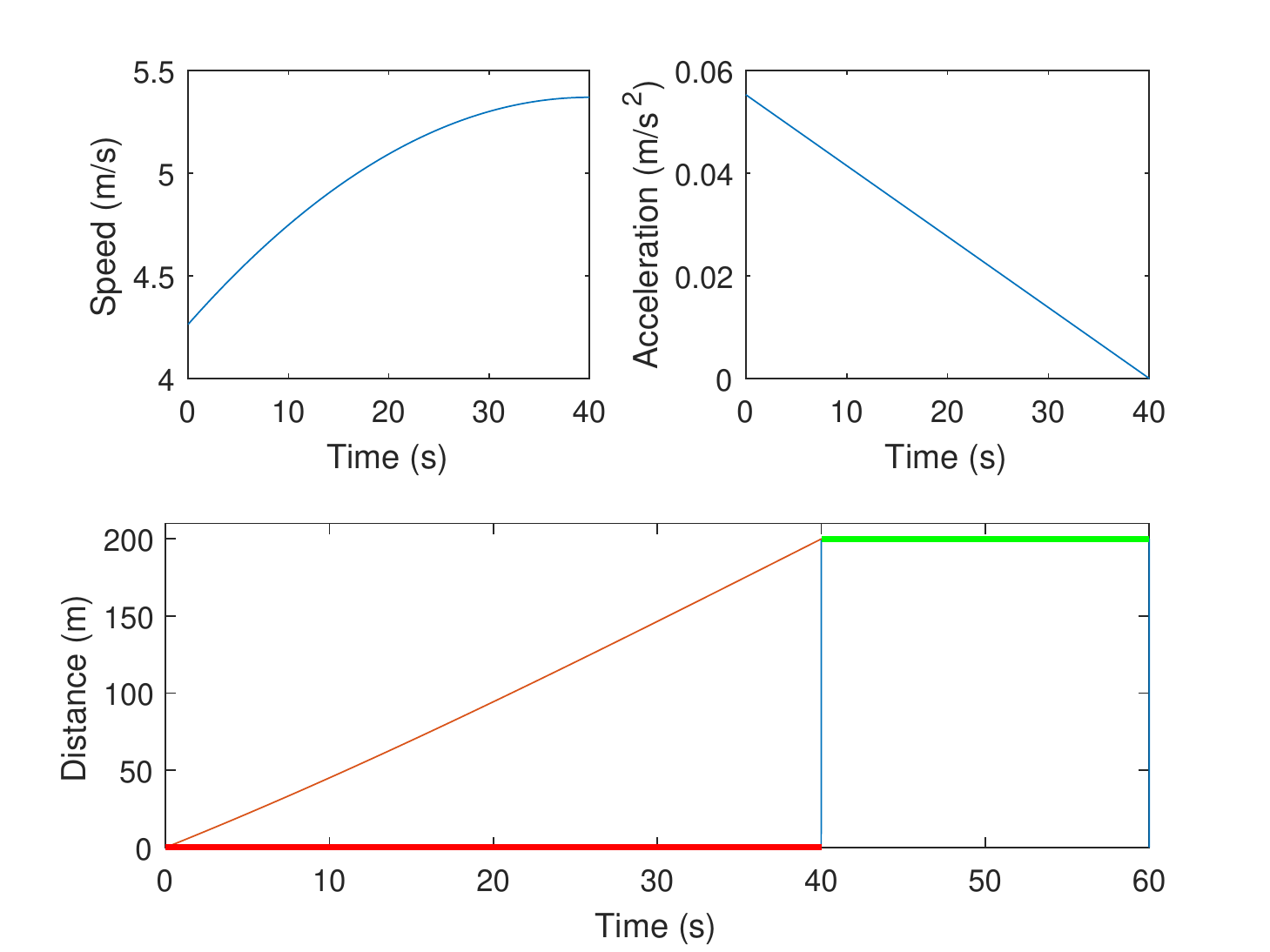}\caption{Case V in Table~\ref{ta2}
with $v_{0}=4.2634$.}%
\label{fig4}%
\end{figure}

In the last two figures, the traffic light starts at a green state. The
following two figures show the case when the traffic light starts at a red
state. It can be inferred from the first two plots that the arrival time
obtained from the free terminal time optimal control problem should be within
the red light interval. Figure~\ref{fig4} shows a case when the initial speed
is slow. The optimal arrival time obtained from the free terminal time optimal
control is 12.1860 seconds. However, the traffic light in the first 40 seconds
is red. The optimal time for the vehicle to arrive at the intersection is 40
seconds. The vehicle has adequate time to accelerate, therefore, it does not
start with full acceleration, and it is unnecessary to accelerate to the
maximum speed. 

\begin{figure}[t]
\centering
\includegraphics[width=\linewidth]{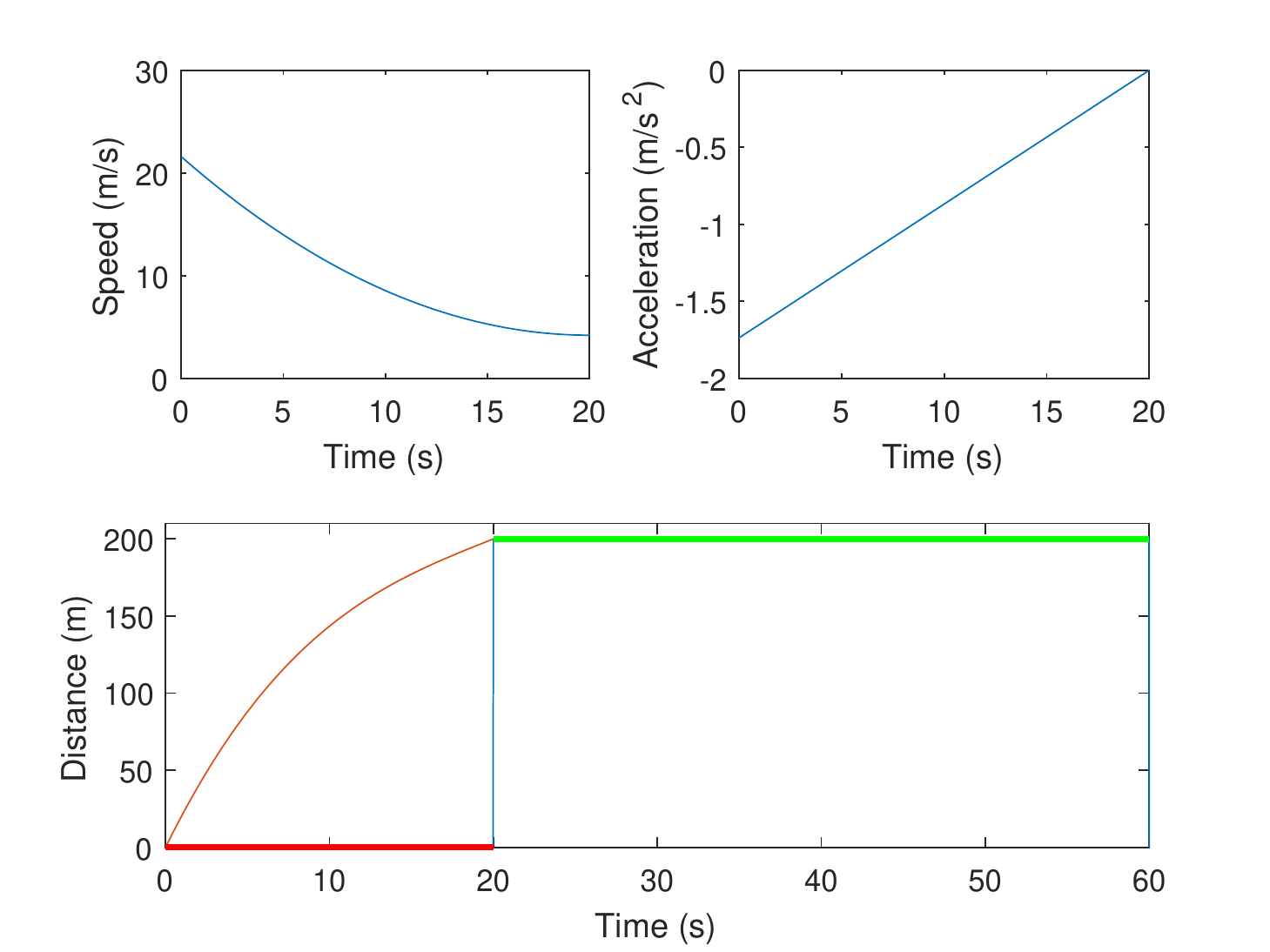}\caption{Case X in Table~\ref{ta3}
with $v_{0}=21.5791$.}%
\label{fig3}%
\end{figure}

Figure~\ref{fig3} exhibits a different traffic light pattern, where the
traffic light in the first 20 seconds is red. Due to a relatively large
initial speed, the vehicle has to decelerate to cross the intersection when
the traffic light is green. 
%

\begin{figure}[t]
\centering
\includegraphics[width=\linewidth]{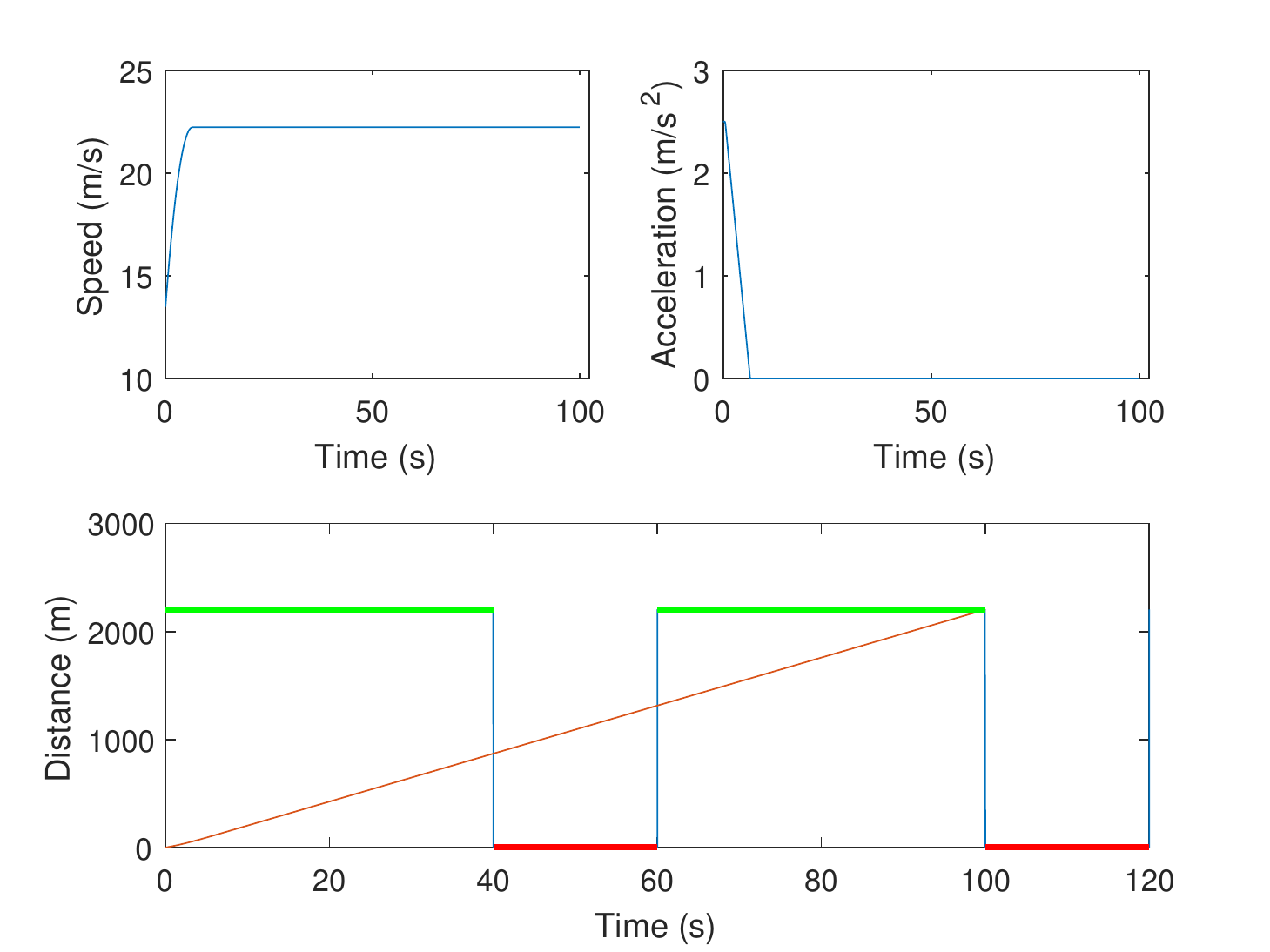}\caption{Case II in
Table~\ref{ta2} with $v_{0}=13.4875$.}%
\label{fig6}%
\end{figure}

In the following, we test the optimal controller on a road of length $2203~m$.
Due to this length, the optimal arrival time usually does not fall within the
first green light cycle, and sometimes it is impossible for the vehicle to
arrive at the traffic light within this cycle. For the case in Fig.~\ref{fig6}%
, the optimal arrival time calculated from the free terminal time optimal
control problem is 102.3476 seconds. Unfortunately, this arrival time belongs
to a red light interval. Therefore, full acceleration is used to reach the
speed limit and cross the intersection at 100 seconds when the traffic light
is green. 

\begin{figure}[t]
\centering
\includegraphics[width=\linewidth]{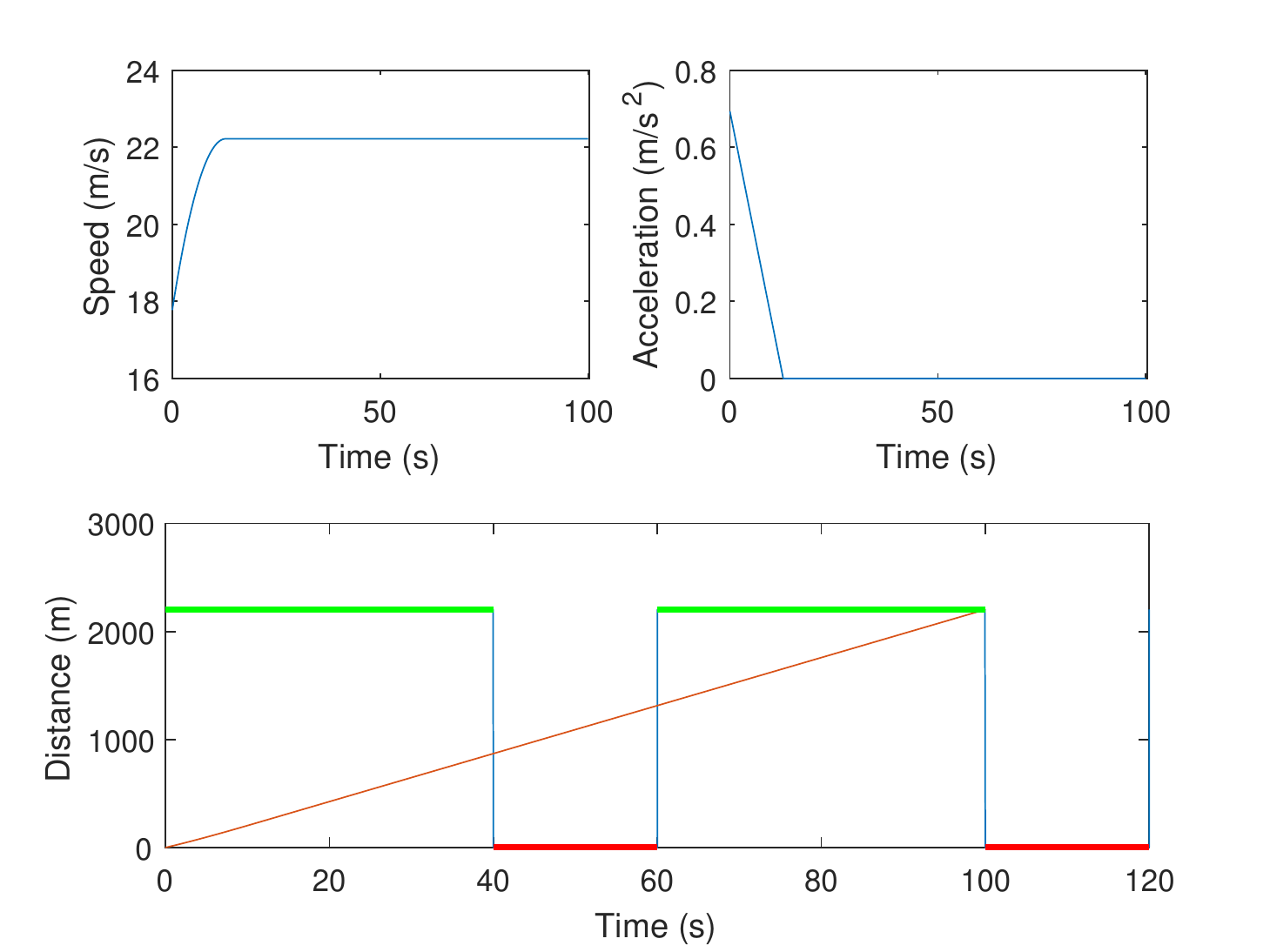}\caption{Case IV in
Table~\ref{ta2} with $v_{0}=17.7745$.}%
\label{fig7}%
\end{figure}

Figure~\ref{fig7} shows the case when the vehicle has a relatively fast
initial speed compared to Fig.~\ref{fig6}. Therefore, the vehicle does not
start with full acceleration to reach the speed limit and catch the green
light at 100 seconds. 

\begin{figure}[t]
\centering
\includegraphics[width=\linewidth]{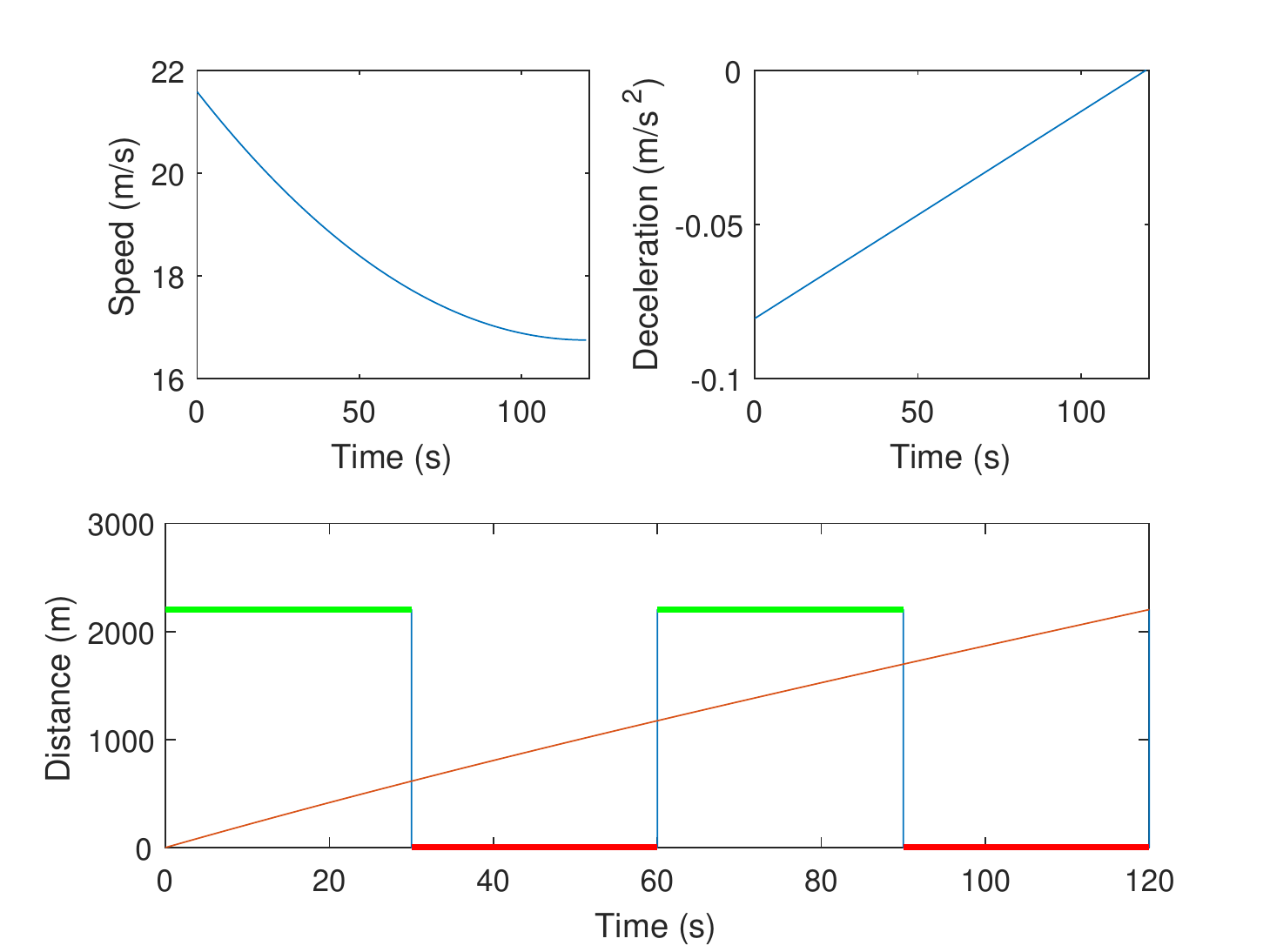}\caption{Case X in Table~\ref{ta3}
with $v_{0}=21.5791$.}%
\label{fig8}%
\end{figure}

For the last case in Fig.~\ref{fig8}, the initial speed is very large. The
best option is to decelerate the vehicle to cross the intersection at 120
seconds when the traffic light is green. 

\textbf{Exploring the time-energy tradeoff. }In order to compare the
performance between $(i)$ autonomous vehicles under the optimal control
developed and $(ii)$ a human driver, we arbitrarily define the following rules
as the driving behavior of a human driver:

\begin{itemize}
\item Full acceleration when the traffic light is green;

\item No acceleration/deceleration when the traffic light is red.
\end{itemize}

We calculate the performance of both autonomous vehicles and human drivers for
the different scenarios encountered from Fig.~\ref{fig1} to Fig.~\ref{fig8},
and summarize the results in Table~\ref{ta4}. The improvement is more than
10\% for the case in Fig.~\ref{fig4}. The performance improvement is
calculated as the performance difference between the human driver and
autonomous vehicle divided by the performance of the human driver. It is
particularly challenging for a human driver to make a decision when he/she
faces a steady red traffic light. \begin{table}[ptb]
\caption{Performance Comparison Between Human Driver (HD) and Autonomous
Vehicle (AV)}%
\label{ta4}
\begin{center}
\renewcommand{\arraystretch}{2.0}
\begin{tabular}
[c]{c|c|c|c}\hline
& HD & AV & Improvement\\\hline
Fig.~\ref{fig1} & 0.1611 & 0.1574 & 2.3\%\\
Fig.~\ref{fig2} & 0.1294 & 0.1263 & 2.4\%\\
Fig.~\ref{fig4} & 0.5965 & 0.5310 & 10.98\%\\
Fig.~\ref{fig3} & 0.2655 & 0.2841 & NA
\tablefootnote{In this case, the human driver approaches the intersection at red light with the speed $21.5791$. We assume that the human driver is able to stop before the traffic light immediately. In addition, we did not consider the energy consumptions of sudden braking and restarting the vehicle.}\\
Fig.~\ref{fig6} & 0.1300 & 0.1224 & 5.85\%\\
Fig.~\ref{fig7} & 0.1406 & 0.1350 & 3.98\%\\
Fig.~\ref{fig8} & 0.1461 & 0.1448 & 0.89\%
\tablefootnote{In this case, the human driver approaches the intersection with the maximum speed at red light. We assume that the human driver is able to stop before the traffic light immediately from the maximum speed. In addition, we did not consider the energy consumptions of sudden braking and restarting the vehicle.}\\\hline
\end{tabular}
\end{center}
\end{table}Also note that the weighting parameter $\rho$ is chosen to be in
favor of travel time rather than energy efficiency. Therefore, the performance
improvement would be larger when we decrease the weighting parameter $\rho$,
which provides a trade-off between energy consumption and travel time.

\begin{figure}[t]
\centering
\includegraphics[width=\linewidth]{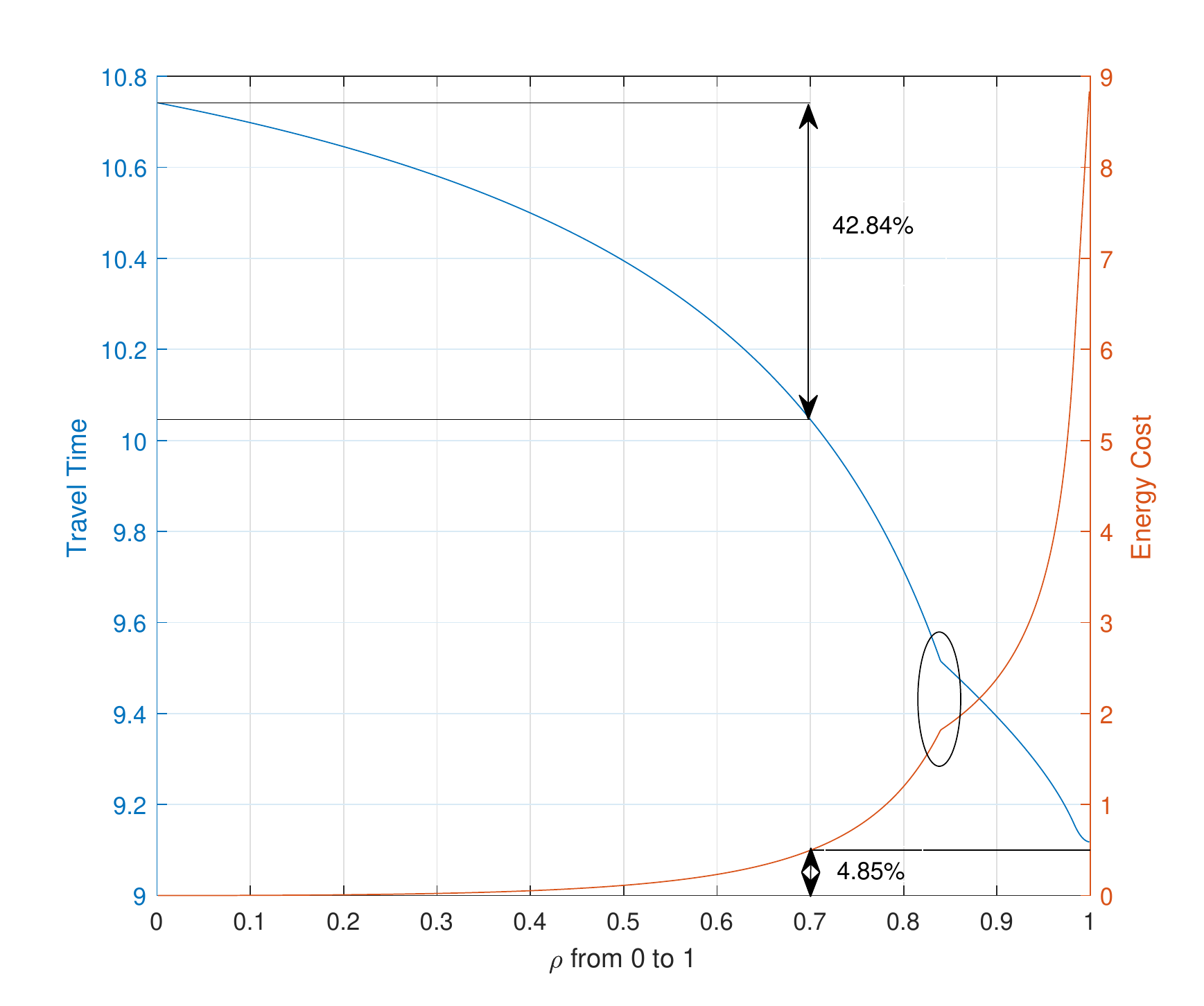}\caption{Trade-off between travel
time and energy consumption}%
\label{fig9}%
\end{figure}

Figure~\ref{fig9} shows the travel time and the energy consumption when we
vary the parameter $\rho$ from 0 to 1. The initial speed is chosen as
$v_{0}=18.6182$. By exploring the trade-off curve, one may select am
appropriate weight parameter $\rho$ depending on a particular application of
interest. For instance, if energy efficiency is a major concern,
Fig.~\ref{fig9} suggests to not select a large value for $\rho$ since the
energy consumption grows rapidly as $\rho$ approaches $1$. On the other hand,
a small $\rho$ is likely not a better option, since we can see that energy
consumption does not significantly increase with $\rho$ increasing as long as
$\rho<0.7$ (approximately). In fact, when $\rho$ increases from 0 to 0.7, the
travel time is significantly reduced by $42.84\%$ whereas the energy
consumption increases by only $4.85\%$. It is noteworthy that both curves show
different trends around the circled area shown in Fig.~\ref{fig9}: this is
mainly because the optimal control has included the full acceleration part
when the parameter $\rho$ is large. 

\section{Conclusions}

\label{con} This paper provided the optimal acceleration/deceleration profile
for autonomous vehicles approaching an intersection based on the traffic light
information, which could be obtained from an intelligent infrastructure via
V2I communication. The solution for the above problem had the key feature of
avoiding idling at a red light. Comparing with similar problems solved by
numerical calculations, we provided a real-time analytical solution. The
proposed algorithm offered better efficiency in terms of travel time and
energy consumption, which has been verified through extensive simulations. The
simulation results showed that the algorithm achieved substantial performance
improvement compared with vehicles with heuristic human driver behavior.

There are a few avenues available for extending this work. In particular,
there is a need to consider a practical scenario where interferences from
other road users present. A possible way of doing this is to predict the
driving behavior of vehicles ahead. It is also desirable to develop a more
general algorithm by taking into account traffic light information at multiple intersections.

\appendices

\section{Proof of Lemma~\ref{L1}}

\label{a10} Let us first consider the case of constant control input. By
solving the differential equation (\ref{dc2}), it is straightforward to get
the expression for $v\left(  t_{1}\right)  $. As a byproduct, we have
$v\left(  t\right)  =v_{0}+\left(  t-t_{0}\right)  u$. By solving the
differential equation (\ref{dc1}), it follows that%
\begin{align*}
x\left(  t_{1}\right)    & =x_{0}+\int_{t_{0}}^{t_{1}}\left[  v_{0}+\left(
t-t_{0}\right)  u\right]  dt\\
& =x_{0}+v_{0}\left(  t_{1}-t_{0}\right)  +\frac{1}{2}u\left(  t_{1}%
-t_{0}\right)  ^{2}\text{.}%
\end{align*}
The energy consumption $J^{u}$ is then easy to obtain.

Next, let us consider the case that $u\left(  t\right)  =u\left(
t_{1}-t\right)  $ for $t\in\left[  t_{0},t_{1}\right]  $. Solving the
differential equation (\ref{dc2}), we obtain%
\begin{align*}
v\left(  t_{1}\right)    & =v_{0}+\int_{t_{0}}^{t_{1}}\left[  u\left(
t_{1}-t\right)  \right]  dt\\
& =v_{0}+\frac{1}{2}u\left(  t_{1}-t_{0}\right)  ^{2}.
\end{align*}
As a byproduct, we have $v\left(  t\right)  =v_{0}+\frac{1}{2}u\left(
t-t_{0}\right)  ^{2}$. Solving the differential equation (\ref{dc1}) yields%
\begin{align*}
x\left(  t_{1}\right)    & =x_{0}+\int_{t_{0}}^{t_{1}}\left[  v_{0}+\frac
{1}{2}u\left(  t-t_{0}\right)  ^{2}\right]  dt\\
& =x_{0}+v_{0}\left(  t_{1}-t_{0}\right)  +\frac{1}{6}u\left(  t_{1}%
-t_{0}\right)  ^{3}\text{.}%
\end{align*}
The energy consumption is then calculated as%
\[
J^{u}=\int_{t_{0}}^{t_{1}}u^{2}\left(  t_{1}-t\right)  ^{2}dt=\frac{1}{3}%
u^{2}\left(  t_{1}-t_{0}\right)  ^{3}.
\]

\section{Proof of Lemma~\ref{l2}}

\label{a1} We will prove the result by a contradiction argument. Let us assume
that $u^{\ast}\left(  t\right)  $ and $t_{p}^{\ast}$ are the optimal control
and the optimal arrival time of Problem~\ref{P2}, respectively. In addition,
we assume that there exists an interval $\left[  t_{1},t_{2}\right]  $ such
that $u^{\ast}(t)<0$. Next, we construct another control input $u\left(
t\right)  $ such that $u\left(  t\right)  =u^{\ast}\left(  t\right)  $ for
$t<t_{1}$, and $u\left(  t\right)  =0$ for $t\in\left[  t_{1},t_{2}\right]  $.
It is then straightforward to get%
\[
v^{\ast}(t_{1})=v(t_{1}),\text{ and }x^{\ast}(t_{1})=x(t_{1}).
\]
We now invoke the comparison lemma \cite{khalil2002nonlinear} which compares
the solutions of the differential inequality $\dot{v}(t)\leq f(t,v)$ with the
solution of the differential equation $\dot{u}(t)=f(t,u)$ and asserts that If
$v_{0}\leq u_{0}$, then $v(t)\leq u(t)$. By applying the comparison principle
to the dynamics of $v(t)$, it follows that%
\begin{equation}
v^{\ast}(t)<v(t)\label{CompLemma}%
\end{equation}
for $t>t_{1}$ until $v^{\ast}(t)=v_{\max}$. By applying the comparison
principle again to the dynamics of $x(t)$, it follows that $x^{\ast}(t)<x(t)$
for $t>t_{1}$. Then, according to the terminal condition%
\[
x^{\ast}(t_{p})<x(t_{p})=l,
\]
we conclude that $t_{p}^{\ast}>t_{p}$, therefore we have%
\begin{equation}
t_{p}-t_{0}<t_{p}^{\ast}-t_{0}.\label{ov}%
\end{equation}
Let $\tau$ be the time when $v(\tau)=v_{\max}$, and we assume that $\tau
>t_{2}$ without loss of generality. The remaining control input of $u(t)$ is
thus constructed as%
\[
u\left(  t\right)  =\left\{
\begin{array}
[c]{ll}%
u^{\ast}\left(  t\right)   & \text{for }t_{2}\leq t<\min\left\{  \tau
,t_{p}\right\}  \\
0 & \text{for }t\geq\min\left\{  \tau,t_{p}\right\}  .
\end{array}
\right.
\]
By using the inequality (\ref{CompLemma}) at $t=\tau$ we have%
\[
v^{\ast}(\tau)<v(\tau)=v_{\max}%
\]
Recalling that $u^{\ast}(t)<0$, $u\left(  t\right)  =0$ for $t\in\left[
t_{1},t_{2}\right]  $, it follows that
\begin{align*}
\int_{t_{0}}^{t_{p}}u^{2}\left(  t\right)  dt= &  \int_{t_{0}}^{t_{1}}%
u^{2}\left(  t\right)  dt+\int_{t_{2}}^{\min\left\{  \tau,t_{p}\right\}
}u^{2}\left(  t\right)  dt\\
< &  \int_{t_{0}}^{t_{1}}u^{\ast}\left(  t\right)  ^{2}dt+\int_{t_{1}}^{t_{2}%
}u^{\ast}\left(  t\right)  ^{2}dt\\
&  +\int_{t_{2}}^{\min\left\{  \tau,t_{p}\right\}  }u^{\ast}\left(  t\right)
^{2}dt\\
\leq &  \int_{t_{0}}^{t_{p}^{\ast}}u^{\ast}\left(  t\right)  ^{2}dt.
\end{align*}
The above inequality together with (\ref{ov}) contradicts the optimality of
$u^{\ast}(t)$ and $t_{p}^{\ast}$ in (\ref{ftt1}) and completes the proof by
contradiction. Therefore, we conclude tha $u^{\ast}(t)\geq0$ for all
$t\in\lbrack t_{0},t_{p}^{\ast}]$.

\section{Calculations for Table~\ref{ta1}}

\label{a2} Let us assume that $\rho_{u}\neq0$, and $\rho_{t}\neq0$.

\subsection{Case I: $v^{*}\left(  t_{p}^{\ast}\right)  =v_{\max}$}

Let us first find the time duration $\delta$ such that $u\left(  t\right)  $
decreases from $u_{\max}$ to $0$ while the speed increases from $v$ to the
maximum speed $v_{\max}$ under the optimal control%
\begin{equation}
\dot{u}\left(  t\right)  =-\frac{\rho_{t}}{2\rho_{u}v_{\max}}\text{.}%
\label{ud}%
\end{equation}
Integrating (\ref{ud}) on both sides yields%
\[
0=u_{\max}-\frac{\rho_{t}}{2\rho_{u}v_{\max}}\delta
\]
which can be simplified as%
\[
\delta=2u_{\max}v_{\max}\frac{\rho_{u}}{\rho_{t}}.
\]
According to Lemma~\ref{L1}, we know that%
\[
v_{\max} = v+u_{\max}^{2}v_{\max}\frac{\rho_{u}}{\rho_{t}},
\]
which can be written as%
\[
v=\left(  1-u_{\max}^{2}\frac{\rho_{u}}{\rho_{t}}\right)  v_{\max}%
\]
with the assumption that%
\[
\left(  1-u_{\max}^{2}\frac{\rho_{u}}{\rho_{t}}\right)  v_{\max}\geq v_{\min}.
\]
For the same amount of time, the distance that the vehicle travels is%
\begin{align*}
d =2u_{\max}v_{\max}^{2}\frac{\rho_{u}}{\rho_{t}}-\frac{2}{3}u_{\max}%
^{3}v_{\max}^{2}\frac{\rho_{u}^{2}}{\rho_{t}^{2}}\text{.}%
\end{align*}

According to Theorem~\ref{T1}, the optimal control can be parameterized in
terms of the speed $v(t)$ as%
\[
\left\{
\begin{array}
[c]{ll}%
u^{\ast}\left(  t\right)  =u_{\max} & \text{if }v\left(  t\right)  \leq\left(
1-u_{\max}^{2}\frac{\rho_{u}}{\rho_{t}}\right)  v_{\max}\\
\dot{u}^{\ast}\left(  t\right)  =\frac{-\rho_{t}}{2\rho_{u}v_{\max}} &
\text{if }\left(  1-u_{\max}^{2}\frac{\rho_{u}}{\rho_{t}}\right)  v_{\max}\leq
v\left(  t\right)  \leq v_{\max}\\
u^{\ast}\left(  t\right)  =0 & \text{if }v\left(  t\right)  =v_{\max}%
\end{array}
\right.
\]

There are different cases depending on the relationship between the initial
speed $v_{0}$ and the road length $l$. Remind that the analysis is under the
assumption that%
\[
\left(  1-u_{\max}^{2}\frac{\rho_{u}}{\rho_{t}}\right)  v_{\max}\geq v_{\min}.
\]

\subsubsection{Case I.1 $v_{0}\leq\left(  1-u_{\max}^{2}\frac{\rho_{u}}%
{\rho_{t}}\right)  v_{\max}$}

(The first column in Table~\ref{ta1}). In this case, the vehicle will first
accelerate to $v\left(  t_{1}\right)  =\left(  1-u_{\max}^{2}\frac{\rho_{u}%
}{\rho_{t}}\right)  v_{\max}$ using the maximum acceleration $u_{\max}$. Then
it will travel a distance $d$ to reach $v_{\max}$. At time $t_{1}$, we have%
\[
x\left(  t_{1}\right) =v_{0}\left(  t_{1}-t_{0}\right)  +\frac{1}{2}\left(
t_{1}-t_{0}\right)  ^{2}u_{\max}\text{.}%
\]
It is easy to figure out that%
\[
t_{1}-t_{0}=\frac{\left(  1-u_{\max}^{2}\frac{\rho_{u}}{\rho_{t}}\right)
v_{\max}-v_{0}}{u_{\max}}\text{.}%
\]
To achieve the maximum speed $v_{\max}$, the road length $l$ must satisfy%
\begin{align*}
l  & \geq x(t_{1})+d\\
& =\frac{v_{\max}^{2}-v_{0}^{2}}{2u_{\max}}+u_{\max}v_{\max}^{2}\frac{\rho
_{u}}{\rho_{t}}-\frac{1}{6}u_{\max}^{3}v_{\max}^{2}\frac{\rho_{u}^{2}}%
{\rho_{t}^{2}}.
\end{align*}

\subsubsection{Case I.2 $\left(  1-u_{\max}^{2}\frac{\rho_{u}}{\rho_{t}%
}\right)  v_{\max}< v_{0}\leq v_{\max}$}

(The third column in Table~\ref{ta1}). In this case, the vehicle will not
start with full acceleration, and we have%
\[
u^{\ast}\left(  t\right)  =\frac{\rho_{t}}{2\rho_{u}}\frac{\tau-t}{v_{\max}},
\]
where $\tau$ is the time when $v\left(  \tau\right)  =v_{\max}$.

According to Lemma~\ref{L1}, we can obtain%
\begin{align}
x\left(  \tau\right)   & =v_{0}\left(  \tau-t_{0}\right)  +\frac{\rho_{t}%
}{6\rho_{u}v_{\max}}\left(  \tau-t_{0}\right)  ^{3},\nonumber
\end{align}
and%
\begin{equation}
v_{\max}=v_{0}+\frac{\rho_{t}}{4\rho_{u}v_{\max}}\left(  \tau-t_{0}\right)
^{2}.\label{ts}%
\end{equation}

We can calculate from (\ref{ts}) to get%
\begin{equation}
\tau-t_{0}=2\sqrt{\left(  v_{\max}-v_{0}\right)  v_{\max}\frac{\rho_{u}}%
{\rho_{t}}}.\label{fc}%
\end{equation}
By using (\ref{fc}), a necessary condition for $v\left(  t\right)  $ to reach
the maximum speed $v_{\max}$ is%
\begin{align}
l  & \geq2v_{0}\sqrt{\left(  v_{\max}-v_{0}\right)  v_{\max}\frac{\rho_{u}%
}{\rho_{t}}}\nonumber\\
& +\frac{4}{3}\left(  v_{\max}-v_{0}\right)  \sqrt{\left(  v_{\max}%
-v_{0}\right)  v_{\max}\frac{\rho_{u}}{\rho_{t}}}.\label{ld}%
\end{align}


\subsection{Case II: $v^{*}\left(  t_{p}^{\ast}\right)  <v_{\max}$}

\subsubsection{Case II.1 $v_{0}<\left(  1-u_{\max}^{2}\frac{\rho_{u}}{\rho
_{t}}\right)  v_{\max}$}

(The second column in Table~\ref{ta1}). In this case, the road length
\[
l<\frac{v_{\max}^{2}-v_{0}^{2}}{2u_{\max}}+u_{\max}v_{\max}^{2}\frac{\rho_{u}%
}{\rho_{t}}-\frac{1}{6}u_{\max}^{3}v_{\max}^{2}\frac{\rho_{u}^{2}}{\rho
_{t}^{2}}%
\]
is not long enough for the vehicle to reach the maximum speed. Let us assume
that the speed when the acceleration starts to decrease at time $t_{1}$ is $v
$. According to Lemma~\ref{L1}, it takes the time%
\[
t _{1}-t_{0}=\frac{v-v_{0}}{u_{\max}}%
\]
for the vehicle to reach the speed $v$ by using the maximum acceleration, and%
\[
x\left(  t_{1}\right) =\frac{ v^{2}-v_{0}^{2}}{2u_{\max}}\text{.}%
\]
The speed $v$ increases to $v^{\ast}\left(  t_{p}^{\ast}\right)  $ by using a
linearly decreasing optimal control from $u_{\max}$ to $0$. It is easy to get
that%
\[
\dot{u}\left(  t\right)  =-\frac{\rho_{t}}{2\rho_{u}v^{\ast}\left(
t_{p}^{\ast}\right)  }.
\]
Therefore, the time for $u_{\max}$ to decrease to $0$ is%
\[
\delta_{2}=2u_{\max}v^{\ast}\left(  t_{p}^{\ast}\right)  \frac{\rho_{u}}%
{\rho_{t}}\text{.}%
\]
According to Lemma~\ref{L1}, we can obtain%
\begin{align*}
v^{\ast}\left(  t_{p}^{\ast}\right)   & =v+u_{\max}^{2}v^{\ast}\left(
t_{p}^{\ast}\right)  \frac{\rho_{u}}{\rho_{t}},
\end{align*}
which is%
\[
v^{\ast}\left(  t_{p}^{\ast}\right)  =\frac{v}{1-\frac{\rho_{u}}{\rho_{t}%
}u_{\max}^{2}}.
\]

By the road length constraint, we are able to calculate $v$ from the equality%
\begin{align*}
l=\frac{ v^{2}-v_{0}^{2}}{2u_{\max}}+v^{2}\frac{2u_{\max}}{1-\frac{\rho_{u}%
}{\rho_{t}}u_{\max}^{2}}\frac{\rho_{u}}{\rho_{t}}+\frac{4}{3}v^{2}\frac
{\rho_{u}^{2}}{\rho_{t}^{2}}\frac{u_{\max}^{3}}{\left(  1-\frac{\rho_{u}}%
{\rho_{t}}u_{\max}^{2}\right)  ^{2}},
\end{align*}
that is,%
\[
v=\sqrt{\frac{2u_{\max}l+v_{0}^{2}}{1+\frac{4u_{\max}^{2}}{1-\frac{\rho_{u}%
}{\rho_{t}}u_{\max}^{2}}\frac{\rho_{u}}{\rho_{t}}+\frac{8}{3}\frac{u_{\max
}^{4}}{\left(  1-\frac{\rho_{u}}{\rho_{t}}u_{\max}^{2}\right)  ^{2}}\frac
{\rho_{u}^{2}}{\rho_{t}^{2}}}}%
\]


\subsubsection{Case II.2 $v_{0}>\left(  1-u_{\max}^{2}\frac{\rho_{u}}{\rho
_{t}}\right)  v_{\max}$}

(The fourth column in Table~\ref{ta1}). In this case, the road length%
\begin{align*}
l< & \,2v_{0}\sqrt{\left(  v_{\max}-v_{0}\right)  v_{\max}\frac{\rho_{u}}%
{\rho_{t}}}\\
& \quad+\frac{4}{3}\left(  v_{\max}-v_{0}\right)  \sqrt{\left(  v_{\max}%
-v_{0}\right)  v_{\max}\frac{\rho_{u}}{\rho_{t}}}%
\end{align*}
is not large enough for the vehicle to reach the speed limit, and the maximum
acceleration $u_{\max}$ will not be used either. According to Theorem~\ref{T1}%
, the optimal control can be parameterized as%
\[
u^{\ast}\left(  t\right)  =\frac{\rho_{t}\left(  t_{p}^{\ast}-t\right)
}{2\rho_{u}v^{\ast}\left(  t_{p}^{\ast}\right)  }.
\]
According to Lemma~\ref{L1}, we have%
\begin{equation}
v^{\ast}\left(  t_{p}^{\ast}\right)  =v_{0}+\frac{\rho_{t}\left(  t_{p}^{\ast
}-t_{0}\right)  ^{2}}{4\rho_{u}v^{\ast}\left(  t_{p}^{\ast}\right)
},\label{eqa1}%
\end{equation}
and%
\begin{equation}
l=v_{0}\left(  t_{p}^{\ast}-t_{0}\right)  +\frac{\rho_{t}}{6\rho_{u}v^{\ast
}\left(  t_{p}^{\ast}\right)  }\left(  t_{p}^{\ast}-t_{0}\right)
^{3}.\label{eqa2}%
\end{equation}
By solving the equation~(\ref{eqa1}), we can obtain%
\[
v^{*}(t_{p}^{*})=\frac{v_{0}+\sqrt{v_{0}^{2}+\frac{\rho_{t}}{\rho_{u}}\left(
t_{p}^{\ast}-t_{0}\right)  ^{2}}}{2}.
\]
By substituting for $v^{*}(t_{p}^{*})$, we are able to obtain $t_{p}^{*}$ from
(\ref{eqa2}).

\section{Detailed Calculations for Table~\ref{ta2}}

\label{a3}

There are different cases depending on the initial speed $v_{0}$, the time
duration $kT+DT$, and the road length $l$.

\subsection{Case I: $u^{*}_{0}=u_{\max}$ and $\dot{u}^{*}\left(  t\right)
=0$}

This case corresponds to $h(v_{0})=0$. The vehicle accelerates fully until it
arrives at the traffic light or the maximum speed is reached. According to
Lemma 1, the vehicle reaches the maximum speed by spending time%
\[
\delta=\frac{v_{\max}-v_{0}}{u_{\max}}.
\]
Depending on the values of $kT+DT$ and $\delta$, we have different energy
consumptions%
\[
J_{1}^{u}=\left\{
\begin{array}
[c]{cc}%
u_{\max}\left(  v_{\max}-v_{0}\right)  & \text{if }\frac{v_{\max}-v_{0}%
}{u_{\max}}<kT+DT\\
u_{\max}^{2}\left(  kT+DT-t_{0}\right)  & \text{if }\frac{v_{\max}-v_{0}%
}{u_{\max}}\geq kT+DT
\end{array}
\right.
\]

For all other cases, $h\left(  v_{0}\right)  >0$, and $\dot{u}\left(
t\right)  \neq0$ for some $t$.

\subsection{Case II: $u^{\ast}\left(  t_{0}\right)  =u_{\max}$, and $v^{\ast
}\left(  t_{p}\right)  =v_{\max}$}

The time $t_{1}$ is when the acceleration starts to decrease, that is,%
\begin{equation}
\frac{1}{2}\frac{u_{\max}^{2}\left(  \tau-t_{1}\right)  }{v_{0}-v_{\max
}+\left(  \tau-t_{0}\right)  u_{\max}}=u_{\max}.\label{eq21}%
\end{equation}
From (\ref{eq21}), we can obtain%
\[
\tau=2\frac{v_{\max}-v_{0}}{u_{\max}}-t_{1}+2t_{0}\text{.}%
\]
According to Lemma \ref{L1},%
\begin{align*}
v\left(  t_{1}\right)   & =v_{0}+u_{\max}\left(  t_{1}-t_{0}\right) \\
x\left(  t_{1}\right)   & =v_{0}\left(  t_{1}-t_{0}\right)  +\frac{1}%
{2}u_{\max}\left(  t_{1}-t_{0}\right)  ^{2}%
\end{align*}
and%
\[
x\left(  \tau\right)  =x\left(  t_{1}\right)  +v\left(  t_{1}\right)  \left(
\tau-t_{1}\right)  +\frac{1}{6}\frac{u_{\max}^{2}\left(  \tau-t_{1}\right)
^{3}}{v_{0}-v_{\max}+\left(  \tau-t_{0}\right)  u_{\max}}.
\]
Therefore, we have%
\begin{equation}
l=\left(  t_{p}-\tau\right)  v_{\max}+x\left(  \tau\right)  .\label{eq22}%
\end{equation}
We can solve the equation (\ref{eq22}) to get $t_{1}$. The energy consumption
can be expressed as%
\[
J_{2}^{u}=\left(  t_{1}-t_{0}\right)  u_{\max}^{2}+\frac{1}{12}\frac{u_{\max
}^{4}\left(  \tau-t_{1}\right)  ^{3}}{\left[  v_{0}-v_{\max}+\left(
\tau-t_{0}\right)  u_{\max}\right]  ^{2}}.
\]

\subsection{Case III: $u^{\ast}\left(  t_{0}\right)  =u_{\max}$, and $v^{\ast
}\left(  t_{p}\right)  <v_{\max}$}

In this case, $\tau=t_{p}$. First, we need to find the time $t_{1}$ such that
the acceleration starts to decrease, that is,%
\[
\frac{1}{2}\frac{u_{\max}^{2}\left(  t_{p}-t_{1}\right)  }{v_{0}-v^{\ast
}\left(  t_{p}\right)  +\left(  t_{p}-t_{0}\right)  u_{\max}}=u_{\max}\text{.}%
\]
By solving the above equation for $v^{\ast}\left(  t_{p}\right) $, we can
obtain%
\begin{equation}
v^{\ast}\left(  t_{p}\right)  =v_{0}+\frac{t_{p}+t_{1}-2t_{0}}{2}u_{\max
}\text{.}\label{eq11}%
\end{equation}

According to Lemma \ref{L1}, the speed and the distance of the vehicle at
$t_{1}$ are%
\[
v^{*}\left(  t_{1}\right)  =v_{0}+\left(  t_{1}-t_{0}\right)  u_{\max},
\]
and%
\[
x^{*}\left(  t_{1}\right)  =v_{0}\left(  t_{1}-t_{0}\right)  +\frac{1}%
{2}u_{\max}\left(  t_{1}-t_{0}\right)  ^{2},
\]
respectively. From the road length constraint%
\begin{equation}
l=x^{*}\left(  t_{1}\right)  +v^{*}\left(  t_{1}\right)  \left(  t_{p}%
-t_{1}\right)  +\frac{1}{3}u_{\max}\left(  t_{p}-t_{1}\right) ^{2}
,\label{eq12}%
\end{equation}
we are able to calculate $t_{1}$. The energy consumption for this case can be
expressed as%
\[
J_{3}^{u}=\frac{u_{\max}^{2}\left(  t_{p}+2t_{1}-3t_{0}\right)  }{3}\text{.}%
\]

\subsection{Case IV $u_{0}^{\ast}<u_{\max}$ and $v^{\ast}\left(  t_{p}\right)
=v_{\max}$}

In this case, the vehicle reaches the maximum speed at $\tau$. According to
Lemma \ref{L1}, we have%
\begin{equation}
v_{\max}=v_{0}+\frac{1}{4}\frac{u^{\ast}\left(  t_{0}\right)  ^{2}\left(
\tau-t_{0}\right)  ^{2}}{v_{0}-v_{\max}+\left(  \tau-t_{0}\right)  u^{\ast
}\left(  t_{0}\right)  }\text{,}\label{eq41}%
\end{equation}
Solving the above equation for $u^{\ast}\left(  t_{0}\right) $ yields%
\[
u^{\ast}\left(  t_{0}\right)  =2\frac{v_{\max}-v_{0}}{\tau-t_{0}}.
\]
With the expression of $u^{\ast}\left(  t_{0}\right)  $ and Lemma \ref{L1}, we
can obtain%
\begin{equation}
l=\frac{1}{3}\left(  v_{0}+2v_{\max}\right)  \left(  \tau-t_{0}\right)
+\left(  t_{p}-\tau\right)  v_{\max}\text{.}\label{eq42}%
\end{equation}
We can calculate $\tau$ from (\ref{eq42}) as%
\[
\tau=\frac{3l+\left(  2v_{\max}+v_{0}\right)  t_{0}-3t_{p}v_{\max}}%
{v_{0}-v_{\max}}\text{.}%
\]
The energy consumption in this case is expressed as%
\[
J_{4}^{u}=\frac{4}{3}\frac{\left(  v_{\max}-v_{0}\right)  ^{2}}{\tau-t_{0}%
}\text{.}%
\]

\subsection{Case V: $u_{0}^{\ast}<u_{\max}$ and $v^{\ast}\left(  t_{p}\right)
<v_{\max}$}

In this case, $\tau=t_{f}$. According to Lemma \ref{L1}, the final speed is%
\begin{equation}
v^{\ast}\left(  t_{p}\right)  =v_{0}+\frac{1}{4}\frac{u^{\ast}\left(
t_{0}\right)  ^{2}\left(  t_{p}-t_{0}\right)  ^{2}}{v_{0}-v^{\ast}\left(
t_{p}\right)  +\left(  t_{p}-t_{0}\right)  u^{\ast}\left(  t_{0}\right)
},\label{eq31}%
\end{equation}
From (\ref{eq31}), we can get%
\[
u^{\ast}\left(  t_{0}\right)  =2\frac{v^{\ast}\left(  t_{p}\right)  -v_{0}%
}{t_{p}-t_{0}}\text{.}%
\]
Using the expression of $u^{\ast}\left(  t_{0}\right)  $ and Lemma \ref{L1},
we can obtain%
\begin{equation}
l=v_{0}\left(  t_{p}-t_{0}\right)  +\frac{2}{3}\left(  v^{\ast}\left(
t_{p}\right)  -v_{0}\right)  \left(  t_{p}-t_{0}\right)  \text{.}\label{eq32}%
\end{equation}
Solving the equation (\ref{eq32}), we have%
\[
v^{\ast}\left(  t_{p}\right)  =\frac{3}{2}\frac{l-v_{0}\left(  t_{p}%
-t_{0}\right)  }{t_{p}-t_{0}}+v_{0}\text{.}%
\]
The energy consumption in this case can be expressed as%
\[
J_{5}^{u}=3\frac{\left[  l-v_{0}\left(  t_{p}-t_{0}\right)  \right]  ^{2}%
}{\left(  t_{p}-t_{0}\right)  ^{3}}.
\]

\section{Detailed Calculations for Table~\ref{ta3}}

\label{a4}

\subsection{Case VII: $u\left(  t_{0}\right)  =u_{\min}$ and $v\left(
t_{p}\right)  =v_{\min}$.}

In this case, the vehicle starts with full deceleration $u_{\min}$, and then
at time $t_{1},$ the deceleration linearly increases until it reaches zero at
$t=\tau$. Therefore, at time $t=t_{1}$, we have%
\[
u_{\min}=\frac{1}{2}\frac{u_{\min}^{2}\left(  \tau-t_{1}\right)  }%
{v_{0}+\left(  \tau-t_{0}\right)  u_{\min}-v_{\min}},
\]
that is,%
\begin{equation}
v_{0}-v_{\min}=\frac{2t_{0}-\tau-t_{1}}{2}u_{\min}.\label{eq51}%
\end{equation}

According to Lemma \ref{L1}, the speed and travel distance of the vehicle at
time $t_{1} $ are%
\[
v\left(  t_{1}\right)  =v_{0}+u_{\min}\left(  t_{1}-t_{0}\right)  ,
\]
and
\[
x\left(  t_{1}\right)  =v_{0}\left(  t_{1}-t_{0}\right)  +\frac{1}{2}u_{\min
}\left(  t_{1}-t_{0}\right)  ^{2},
\]
respectively. At time $\tau,$ we have%
\[
x\left(  \tau\right)  =x\left(  t_{1}\right)  +v\left(  t_{1}\right)  \left(
\tau-t_{1}\right)  +\frac{1}{6}\frac{u_{\min}^{2}\left(  \tau-t_{1}\right)
^{3}}{v_{0}+\left(  \tau-t_{0}\right)  u_{\min}-v_{\min}}%
\]
To satisfy the road length constraint, we must have%
\begin{equation}
l=x\left(  \tau\right)  +\left(  kT+T-\tau\right)  v_{\min}\text{.}%
\label{eq52}%
\end{equation}
We can solve (\ref{eq51}) to obtain%
\[
\tau=2t_{0}-t_{1}+2\frac{v_{\min}-v_{0}}{u_{\min}}%
\]
and (\ref{eq52}) to get $t_{1}$. The energy consumption in this case can be
expressed as%
\[
J_{7}^{u}=\left(  t_{1}-t_{0}\right)  u_{\min}^{2}+\frac{1}{12}\frac{u_{\min
}^{4}\left(  \tau-t_{1}\right)  ^{3}}{\left[  v_{0}+\left(  \tau-t_{0}\right)
u_{\min}-v_{\min}\right]  ^{2}}.
\]

\subsection{Case VIII: $u^{\ast}\left(  t_{0}\right)  =u_{\min}$ and $v^{\ast
}\left(  t_{p}\right)  >v_{\min}$.}

In this case, $\tau=t_{f}$. The vehicle starts with full deceleration
$u_{\min}$, and at time $t_{1}$, the deceleration starts to increase.
Similarly, we have%
\begin{equation}
v_{0}-v^{\ast}\left(  t_{p}\right)  =\frac{2t_{0}-t_{p}-t_{1}}{2}u_{\min
}.\label{eq61}%
\end{equation}
According to Lemma \ref{L1}, we know that%
\begin{align*}
v^{\ast}\left(  t_{1}\right)   & =v_{0}+u_{\min}\left(  t_{1}-t_{0}\right) ,\\
x^{\ast}\left(  t_{1}\right)   & =v_{0}\left(  t_{1}-t_{0}\right)  +\frac
{1}{2}u_{\min}\left(  t_{1}-t_{0}\right)  ^{2}.
\end{align*}
Solving (\ref{eq61}), we can get%
\[
v^{\ast}\left(  t_{p}\right)  =v_{0}+\frac{t_{1}+t_{p}-2t_{0}}{2}u_{\min
}\text{.}%
\]
Using the expression of $v^{\ast}\left(  t_{p}\right)  $, we can obtain
$t_{1}$ by solving the following equation%
\begin{equation}
l=x^{\ast}\left(  t_{1}\right)  +v^{\ast}\left(  t_{1}\right)  \left(
t_{p}-t_{1}\right)  +\frac{1}{3}u_{\min}\left(  t_{p}-t_{1}\right)
^{2}.\label{eq62}%
\end{equation}

The energy consumption in this case can be expressed as%
\[
J_{8}^{u}=\frac{u_{\min}^{2}\left(  t_{p}+2t_{1}-3t_{0}\right)  }{3}.
\]

\subsection{Case IX: $u^{\ast}\left(  t_{0}\right)  <u_{\min}$, and $v^{\ast
}\left(  t_{p}\right)  =v_{\min}$.}

In this case, the vehicle starts with linearly increasing deceleration until
it reaches the minimum speed $v_{\min}$.

According to Lemma \ref{L1}, we have%
\begin{equation}
v_{\min}=v_{0}+\frac{1}{4}\frac{u^{\ast}\left(  t_{0}\right)  ^{2}\left(
\tau-t_{0}\right)  ^{2}}{v_{0}+\left(  \tau-t_{0}\right)  u^{\ast}\left(
t_{0}\right)  -v_{\min}}.\label{eq71}%
\end{equation}
Solving $u^{\ast}\left(  t_{0}\right) $ in (\ref{eq71}) yields%
\[
u^{\ast}\left(  t_{0}\right)  =2\frac{v_{\min}-v_{0}}{\tau-t_{0}}\text{.}%
\]
According to Lemma \ref{L1} and the expression of $u^{\ast}\left(
t_{0}\right)  $, the distance of the vehicle at time $\tau$ is given as%
\[
x\left(  \tau\right)  =\frac{1}{3}\left(  2v_{\min}+v_{0}\right)  \left(
\tau-t_{0}\right)  .
\]
Then, we can solve $\tau$ from the following equation%
\begin{equation}
l=x\left(  \tau\right)  +\left(  KT+T-\tau\right)  v_{\min}\text{,}%
\label{eq72}%
\end{equation}
that is,%
\[
\tau=\frac{3l+\left(  2v_{\min}+v_{0}\right)  t_{0}-3t_{p}v_{\min}}%
{v_{0}-v_{\min}}.
\]
The energy consumption in this case can be expressed as%
\[
J_{9}^{u}=\frac{4}{3}\frac{\left(  v_{\min}-v_{0}\right)  ^{2}}{\tau-t_{0}}.
\]

\subsection{Case X: $u^{\ast}\left(  t_{0}\right)  <u_{\min}$, and $v^{\ast
}\left(  t_{p}\right)  <v_{\min}$.}

In this case, $\tau=t_{p}$. The optimal control only contains the linear
increasing deceleration process. According to Lemma \ref{L1}, we have%
\begin{align}
& v^{\ast}\left(  t_{p}\right)  =v_{0}+\frac{1}{4}\frac{u^{\ast}\left(
t_{0}\right)  ^{2}\left(  t_{p}-t_{0}\right)  ^{2}}{v_{0}+\left(  t_{p}%
-t_{0}\right)  u^{\ast}\left(  t_{0}\right)  -v\left(  t_{p}\right)
},\label{eq81}\\
& l =v_{0}\left(  t_{p}-t_{0}\right)  +\frac{1}{6}\frac{u^{\ast}\left(
t_{0}\right)  ^{2}\left(  t_{p}-t_{0}\right)  ^{3}}{v_{0}+\left(  t_{p}%
-t_{0}\right)  u^{\ast}\left(  t_{0}\right)  -v^{\ast}\left(  t_{p}\right)
}.\label{eq82}%
\end{align}
We can solve $u^{\ast}\left(  t_{0}\right)  $ and $v^{\ast}\left(
t_{p}\right)  $ from (\ref{eq81}) and (\ref{eq82}) to obtain%
\[
u^{\ast}\left(  t_{0}\right)  =2\frac{v^{\ast}\left(  t_{p}\right)  -v_{0}%
}{t_{p}-v_{0}},
\]
and%
\[
v^{\ast}\left(  t_{p}\right)  =\frac{3}{2}\frac{l-v_{0}\left(  t_{p}%
-t_{0}\right)  }{t_{p}-t_{0}}+v_{0}.
\]
The energy consumption in this case is%
\[
J_{10}^{u}=3\frac{\left[  l-v_{0}\left(  t_{p}-t_{0}\right)  \right]  ^{2}%
}{\left(  t_{p}-t_{0}\right)  ^{3}}.
\]


\begin{thebibliography}{10}
\providecommand{\url}[1]{#1}
\csname url@samestyle\endcsname
\providecommand{\newblock}{\relax}
\providecommand{\bibinfo}[2]{#2}
\providecommand{\BIBentrySTDinterwordspacing}{\spaceskip=0pt\relax}
\providecommand{\BIBentryALTinterwordstretchfactor}{4}
\providecommand{\BIBentryALTinterwordspacing}{\spaceskip=\fontdimen2\font plus
\BIBentryALTinterwordstretchfactor\fontdimen3\font minus
  \fontdimen4\font\relax}
\providecommand{\BIBforeignlanguage}[2]{{%
\expandafter\ifx\csname l@#1\endcsname\relax
\typeout{** WARNING: IEEEtran.bst: No hyphenation pattern has been}%
\typeout{** loaded for the language `#1'. Using the pattern for}%
\typeout{** the default language instead.}%
\else
\language=\csname l@#1\endcsname
\fi
#2}}
\providecommand{\BIBdecl}{\relax}
\BIBdecl

\bibitem{schrank20152015}
D.~Schrank, B.~Eisele, T.~Lomax, and J.~Bak, ``2015 urban mobility scorecard,''
  Texas A\&M Transportation Institute and INRIX, Tech. Rep., 2015.

\bibitem{li2014survey}
L.~Li, D.~Wen, and D.~Yao, ``A survey of traffic control with vehicular
  communications,'' \emph{IEEE Trans. Intell. Transport. Syst.}, vol.~15,
  no.~1, pp. 425--432, 2014.

\bibitem{gilbert1976vehicle}
E.~G. Gilbert, ``Vehicle cruise: Improved fuel economy by periodic control,''
  \emph{Automatica}, vol.~12, no.~2, pp. 159 -- 166, 1976.

\bibitem{hooker1998optimal}
J.~Hooker, ``Optimal driving for single-vehicle fuel economy,''
  \emph{Transportation Research Part A: General}, vol.~22, no.~3, pp. 183 --
  201, 1988.

\bibitem{hellstrom2010design}
E.~Hellstrom, J.~Aslund, and L.~Nielsen, ``Design of an efficient algorithm for
  fuel-optimal look-ahead control,'' \emph{Control Engineering Practice},
  vol.~18, no.~11, pp. 1318 -- 1327, 2010.

\bibitem{li2012minimum}
S.~E. Li, H.~Peng, K.~Li, and J.~Wang, ``Minimum fuel control strategy in
  automated car-following scenarios,'' \emph{IEEE Transactions on Vehicular
  Technology}, vol.~61, no.~3, pp. 998--1007, 2012.

\bibitem{fleck2016adaptive}
J.~L. Fleck, C.~G. Cassandras, and Y.~Geng, ``Adaptive quasi-dynamic traffic
  light control,'' \emph{IEEE Trans. Control Syst. Technol.}, vol.~24, no.~3,
  pp. 830--842, 2016.

\bibitem{milanes2010controller}
V.~Milanes, J.~Perez, E.~Onieva, and C.~Gonzalez, ``Controller for urban
  intersections based on wireless communications and fuzzy logic,'' \emph{IEEE
  Trans. Intell. Transport. Syst.}, vol.~11, no.~1, pp. 243--248, 2010.

\bibitem{alonso2011autonomous}
J.~Alonso, V.~Milanés, J.~Pérez, E.~Onieva, C.~González, and T.~de~Pedro,
  ``Autonomous vehicle control systems for safe crossroads,''
  \emph{Transportation Research Part C: Emerging Technologies}, vol.~19, no.~6,
  pp. 1095 -- 1110, 2011.

\bibitem{huang2012assessing}
S.~Huang, A.~W. Sadek, and Y.~Zhao, ``Assessing the mobility and environmental
  benefits of reservation-based intelligent intersections using an integrated
  simulator,'' \emph{IEEE Trans. Intell. Transport. Syst.}, vol.~13, no.~3, pp.
  1201--1214, 2012.

\bibitem{kim2014mpc}
K.~D. Kim and P.~R. Kumar, ``An mpc-based approach to provable system-wide
  safety and liveness of autonomous ground traffic,'' \emph{IEEE Trans. Autom.
  Control}, vol.~59, no.~12, pp. 3341--3356, 2014.

\bibitem{zhang2016optimal}
Y.~J. Zhang, A.~A. Malikopoulos, and C.~G. Cassandras, ``Optimal control and
  coordination of connected and automated vehicles at urban traffic
  intersections,'' in \emph{Proc. of the 2016 American Control Conference},
  2016, pp. 6227--6232.

\bibitem{rios2017survey}
J.~Rios-Torres and A.~A. Malikopoulos, ``A survey on the coordination of
  connected and automated vehicles at intersections and merging at highway
  on-ramps,'' \emph{IEEE Trans. Intell. Transport. Syst.}, vol.~18, no.~5, pp.
  1066--1077, 2017.

\bibitem{v2i}
\emph{http://www.audi.com/en/innovation/connect/smart-city.html}.

\bibitem{asadi2011predictive}
B.~Asadi and A.~Vahidi, ``Predictive cruise control: Utilizing upcoming traffic
  signal information for improving fuel economy and reducing trip time,''
  \emph{IEEE Trans. Control Syst. Technol.}, vol.~19, no.~3, pp. 707--714,
  2011.

\bibitem{kamal2013model}
M.~A.~S. Kamal, M.~Mukai, J.~Murata, and T.~Kawabe, ``Model predictive control
  of vehicles on urban roads for improved fuel economy,'' \emph{IEEE
  Transactions on Control Systems Technology}, vol.~21, no.~3, pp. 831--841,
  2013.

\bibitem{mahler2014optimal}
G.~Mahler and A.~Vahidi, ``An optimal velocity-planning scheme for vehicle
  energy efficiency through probabilistic prediction of traffic-signal
  timing,'' \emph{IEEE Trans. Intell. Transport. Syst.}, vol.~15, no.~6, pp.
  2516--2523, 2014.

\bibitem{wan2016optimal}
N.~Wan, A.~Vahidi, and A.~Luckow, ``Optimal speed advisory for connected
  vehicles in arterial roads and the impact on mixed traffic,''
  \emph{Transportation Research Part C: Emerging Technologies}, vol.~69, pp.
  548 -- 563, 2016.

\bibitem{de2016eco}
G.~De~Nunzio, C.~Canudas~de Wit, P.~Moulin, and D.~Di~Domenico, ``Eco-driving
  in urban traffic networks using traffic signals information,''
  \emph{International Journal of Robust and Nonlinear Control}, vol.~26, no.~6,
  pp. 1307--1324, 2016.

\bibitem{hartl1995survey}
R.~F. Hartl, S.~P. Sethi, and R.~G. Vickson, ``A survey of the maximum
  principles for optimal control problems with state constraints,'' \emph{SIAM
  Review}, vol.~37, no.~2, pp. 181--218, 1995.

\bibitem{malikopoulos2008real}
A.~Malikopoulos, ``{Real-Time, Self-Learning Identification and Stochastic
  Optimal Control of Advanced Powertrain Systems},'' Ph.D. dissertation, The
  University of Michigan, 2008.

\bibitem{khalil2002nonlinear}
H.~K. Khalil, \emph{{Nonlinear Systems}}, 3rd~ed.\hskip 1em plus 0.5em minus
  0.4em\relax Prentice Hall, 2002.

\end{thebibliography}
\end{document}